\documentclass[a4paper,12pt]{article}

\pdfoutput=1
\usepackage{amsmath}
\usepackage{amssymb}
\usepackage{amsfonts}
\usepackage{mathrsfs}
\usepackage{bbm}
\usepackage{graphicx,subfigure}

\usepackage{bookmark}

\usepackage{cite}

\usepackage{calc}
\usepackage{hyperref}
\usepackage{array}

\usepackage{ulem}

\usepackage{multirow}
\usepackage{color}
\usepackage{xcolor,colortbl}

\usepackage{psfrag}
\usepackage{pstricks}
\usepackage{epsfig}

\allowdisplaybreaks

\newlength{\dinwidth}
\newlength{\dinmargin}
\definecolor{nicered}{rgb}{1.0,0.0,0.2}

\definecolor{color1}{rgb}{0.9,.4,.2}

\definecolor{color2}{rgb}{0.3,.6,.7}

\definecolor{color3}{rgb}{0.7,.2,.7}

\usepackage{color}

\begin{document}

\title{
\vspace*{-0.5cm}
\bf \Large
Effect of $K^0-\bar{K}^0$ Mixing on CP and CPT Violations in $B_{c}^{\pm}\rightarrow B^{\pm} K_{S,L}^{0}$ Decays}

\author{Xiao-Dong Cheng$^{1}$\footnote{chengxd@mails.ccnu.edu.cn},  Ru-Min Wang$^{2}$\footnote{ruminwang@sina.com}, Xing-Bo Yuan$^{3}$\footnote{y@mail.ccnu.edu.cn}\\
\\
{$^1$\small College of Physics and Electronic Engineering,}\\[-0.2cm]
{    \small Xinyang Normal University, Xinyang 464000, People's Republic of China}\\[-0.1cm]
{$^2$\small College of Physics and Communication Electronics,}\\[-0.2cm]
{    \small JiangXi Normal University, NanChang 330022, People's Republic of China}\\[-0.1cm]
{$^3$\small Institute of Particle Physics and Key Laboratory of Quark and Lepton Physics (MOE),}\\[-0.2cm]
{    \small Central China Normal University, Wuhan 430079, People's Republic of China}\\[-0.1cm]}

\date{}
\maketitle
\bigskip\bigskip
\maketitle
\vspace{-1.2cm}

\begin{abstract}
{\noindent}The large data sample of the $B_c$ meson collected at the LHC experiment and the HL-LHC experiment provides us the opportunity to study the $B_c$ decays and the related physics. In this paper, we investigate the effect of $K^0-\bar{K}^0$ mixing on the the branching ratios, CP violations and CPT violations in the $B_{c}^{\pm}\rightarrow B^{\pm} K_{S,L}^{0}$ decays. We find that some of the $B_c^{\pm}\rightarrow B^{\pm} K_{S,L}^0\rightarrow f_{B^{\pm}} f_{K_{S,L}^0}$ decay chains have large branching ratios, whose maximum value can exceed the order of $10^{-6}$, the minimum number of $B_c^\pm$ events times efficiency for observing the decays at three standard deviations (3$\sigma$) level is about $ 10^6$. We study the CP asymmetries in the $B_c^{\pm}\rightarrow B^{\pm} K_{S,L}^0$ decays and find that the CP asymmetries can exceed the order of $10^{-3}$, which are dominated by $K^0-\bar{K}^0$ mixing. We give the most promising processes to observe the CP violations and the ranges of the numbers of $B_c^\pm$ events-times-efficiency needed to observe the CP asymmetries at a significance of 3$\sigma$ in these decays. We investigate the possibility to constraint the CPT violation parameter $z$ in the $B_c^{\pm}\rightarrow B^{\pm} K_{S,L}^0\rightarrow f_{B^{\pm}} f_{K_{S,L}^0}$ decays and give the most promising processes to constraint the parameter $z$.
\end{abstract}
\newpage

\section{Introduction}
\label{sec:intro}
CP violation provides deep insights into the nature and plays an important role in explaining for our matter-dominated universe~\cite{Sakharov:1967dj,Riotto:1998bt}. In the Standard Model (SM), effects of CP violations in the charm sector, unlike the kaon and B meson systems, are expected to be rather small~\cite{Buccella:1992sg,Buccella:1994nf,Cheng:2012wr}. During the past decade, many theoretical and experimental efforts have so far been made to study CP violations in the D system and charm baryon~\cite{Li:2021uhk,Aaij:2020wil,Aaij:2019vnt,Li:2021iwf,Zhang:2021zhr,Lenz:2020awd,Unal:2020ezc,Saur:2020rgd,Wang:2017gxe,Yu:2017oky,Cheng:2012xb,Azimov:1999gw,Azimov:1999dj,Lipkin:1999qz,Xing:1995jg}. In 2019, the LHCb collaboration reported a first confirmed observation of the CP violations in charm decays via measuring the difference of time-integrated CP asymmetries of $D^0\rightarrow K^+ K^-$ and $D^0\rightarrow {\pi}^+ {\pi}^-$ decays with the result of $(1.54\pm 0.29)\times 10^{-3}$~\cite{Aaij:2019kcg}, the significance of the measured deviation from zero is $5.3\sigma$. However, studies of CP-violating processes in the $B_c$ decays that proceed via the c quark decay with the b quark as a spectator are very scarce.

CPT invariance is one of the most fundamental symmetries in physics and is based on three assumptions: Unitarity, Locality and Lorentz invariance~\cite{Lueders:1992dq}. Therefore, a strong motivation for experimental and theoretical studies on CPT theorem is to test the CPT symmetry~\cite{Karan:2017coa,Anastasi:2018qqf,Karan:2020ada,Karan:2020yhk,Domenico:2020bbk}. The $K^0-{\bar{K}}^0$ mixing system is one of the most intriguing processes to study CP and CPT violations.

The decays with final states including $K_S^0$ or $K_L^0$ can be used to study CP violation~\cite{Amorim:1998pi} and CPT violation. In these decays, $K^0-{\bar{K}}^0$ mixing have a non-negligible effect on CP violation and CPT violation, even play a dominant role. The CP asymmetries in the decays $D^+\rightarrow K_S^0 \pi^+$ and $\tau^+\rightarrow\pi^+ K_S^0 \bar{\nu}_{\tau}$ have been measured by Belle~\cite{Ko:2012pe,Ko:2010ng}, BaBar~\cite{delAmoSanchez:2011zza,BABAR:2011aa}, CLEO~\cite{Mendez:2009aa,Dobbs:2007ab} and FOCUS~\cite{Link:2001zj} collaborations. There exist $2.8\sigma$ discrepancy observed between the BaBar measurement and the SM prediction of the CP asymmetry in the $\tau^+\rightarrow\pi^+K_S \bar{\nu}_{\tau}$ decay~\cite{Grossman:2011zk,Bigi:2012km,Poireau:2012by}. Because the direct CP violation in this decay is absented at the tree level in the SM, the discrepancy could be a hint of the physics beyond the SM, several possible new-physics proposals are put forwarded~\cite{Chen:2020uxi}. However, the result for the CP asymmetry in $\tau^+\rightarrow\pi^+K_S \bar{\nu}_{\tau}$ decay suffers large uncertainty and no unambiguous conclusion can be drawn, so more precise data and more reactions are needed in both experiment and theory.

The $B_c$ meson is the lowest bound state of the doubly heavy-flavored ($\bar{b}c$) system, here both the $b$ and $c$ quark can decay while the other serves as a spectator, so the $B_c$ physics must be very rich~\cite{Brambilla:2010cs,QuarkoniumWorkingGroup:2004kpm,Zheng:2020ult,Cheng:2021svx}. At the LHC experiment, around $5\times 10^{10}$ $B_c$ mesons per year could be produced~\cite{Zhou:2020bnm,Xiao:2011zz}, which provides us the opportunity to get more information of this particle and study the related physics.
In this paper, we investigate the effect of $K^0-\bar{K}^0$ mixing on CP and CPT violations in the $B_{c}^{\pm}\rightarrow B^{\pm} K_{S,L}^{0}$ decays. We predict the branching ratio of these decays and calculate the CP and CPT asymmetries which are dominated by the effect of $K^0-\bar{K}^0$ mixing. In addition, we also explore the sensitivity of these measurements in experiment.
\section{Branching fractions of the $B_{c}^{\pm}\rightarrow B^{\pm} K_{S,L}^{0}$ decays}
\label{sec:decaywidth}
\begin{figure}[t]
\centering
\hspace{0cm}\includegraphics[width=0.36\textwidth]{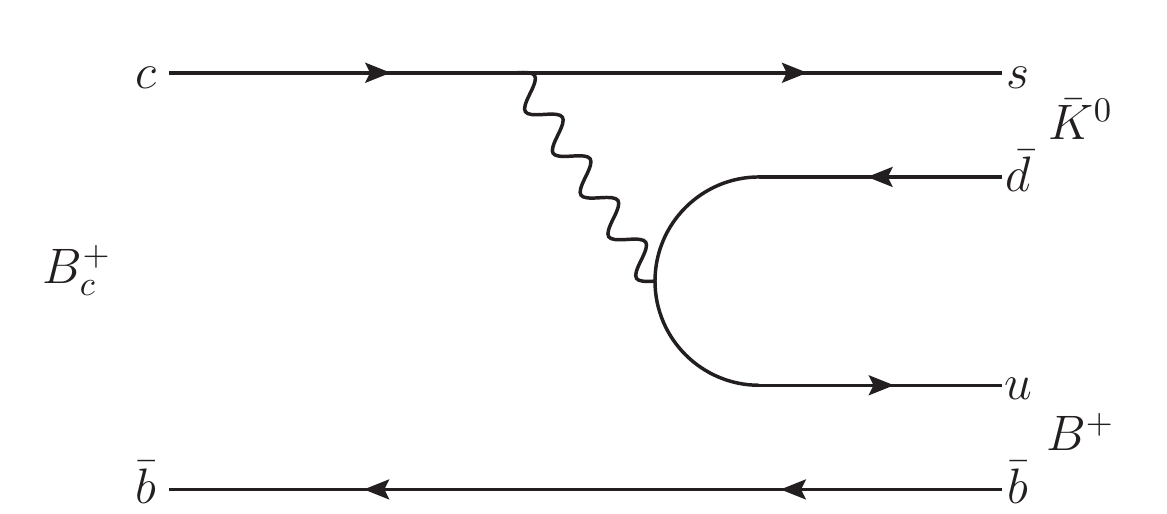}
\hspace{2cm}\includegraphics[width=0.36\textwidth]{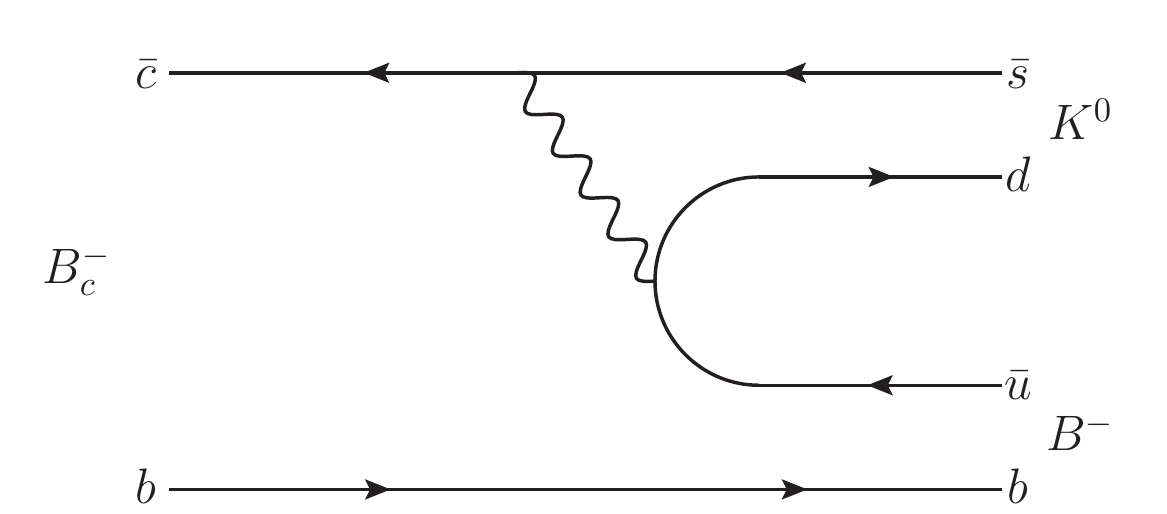}
\caption{\small Quark diagrams for the Cabibbo-allowed $B_c^+\rightarrow B^{+} \bar{K}^{0}$ and $B_c^-\rightarrow B^{-} K^{0}$ decays.}
\label{cabibbofavorbcbukz}
\end{figure}

\begin{figure}[t]
\centering
\hspace{0cm}\includegraphics[width=0.36\textwidth]{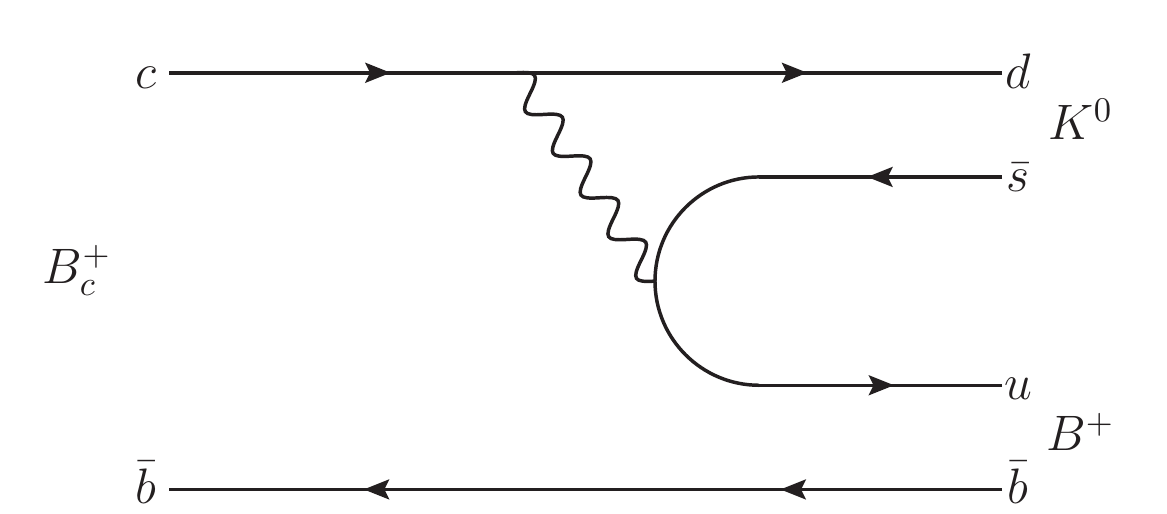}
\hspace{2cm}\includegraphics[width=0.36\textwidth]{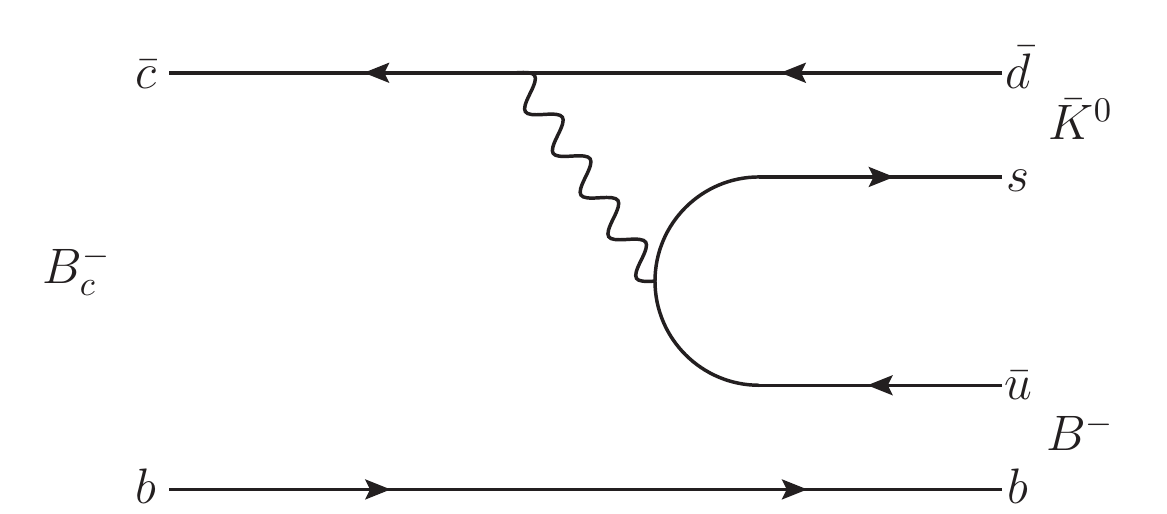}
\caption{\small Quark diagrams for the doubly Cabibbo-suppressed $B_c^+\rightarrow B^{+} K^{0}$ and $B_c^-\rightarrow B^{-} \bar{K}^{0}$ decays.}
\label{dousupcabibbcbukzb}
\end{figure}

Within the SM, the decays $B_c^+\rightarrow B^{+} \bar{K}^{0}$ and its charge conjugate can occur through the Cabibbo-favored channels, which are shown in Fig~\ref{cabibbofavorbcbukz}. The decays $B_c^+\rightarrow B^{+} K^{0}$ and its charge conjugate can proceed via the doubly Cabibbo-suppressed channels, as shown in Fig~\ref{dousupcabibbcbukzb}. The resulting effective Hamiltonian for these decays can be written as
\begin{align}
\mathcal H_{\rm eff}=\frac{G_F}{\sqrt{2}} \left[V_{cs}^{\ast}V_{ud} \bar{s}\gamma^{\mu}(1-\gamma_{5})c\hspace{0.05cm}\bar{u}\gamma_{\mu}(1-\gamma_{5})d+V_{cd}^{\ast}V_{us} \bar{d}\gamma^{\mu}(1-\gamma_{5})c\hspace{0.05cm}\bar{u}\gamma_{\mu}(1-\gamma_{5})s \right]+\text{h.c.}, \label{Eq:effhamilton1}
\end{align}
where $G_F$ is the Fermi coupling constant; $V_{ij}$ denotes the CKM matrix element. To calculate the hadronic matrix elements of the effective Hamiltonian, we work with the factorization assumption~\cite{Fakirov:1977ta,Wang:2008tf,Cheng:2012wr}, then, we need to reconstruct the the effective Hamiltonian in a form suitable for the use of this assumption by Fierz transformation~\cite{Giri:1998qf}. By making use of the Fierz transformation, we obtain the factorized Hamiltonian
\begin{align}
\mathcal H_{\rm eff}^{fac}=-\frac{G_F}{\sqrt{2} N_{c}} \left[V_{cs}^{\ast}V_{ud}\bar{s}\gamma^{\mu}(1-\gamma_{5})d\hspace{0.05cm}\bar{u}\gamma_{\mu}(1-\gamma_{5})c +V_{cd}^{\ast}V_{us}\bar{d}\gamma^{\mu}(1-\gamma_{5})s\hspace{0.05cm}\bar{u}\gamma_{\mu}(1-\gamma_{5})c \right]+\text{h.c.}, \label{Eq:effhamilton2}
\end{align}
with $N_c=3$ is the number of colors.

The decay constant for pseudoscalar meson $K^0$ are defined by~\cite{Branco:1999fs}
\begin{align}
\left \langle K^0 (q_t) \left|\bar{d} \gamma^\mu \gamma_5 s\right|0\right\rangle=iq_t^\mu f_{K}, \label{Eq:kdecaycons}\\
\left \langle |\bar{K}^0 (q_t) \left|\bar{s} \gamma^\mu \gamma_5 d\right|0\right\rangle=iq_t^\mu f_{K}, \label{Eq:kbdecaycons}
\end{align}
here, the CP transformation property ${\mathcal CP}\left|K^0 (q_t^0,\vec{q_t}\hspace{0.02cm})\right\rangle=-\left|K^0 (q_t^0,-\vec{q_t}\hspace{0.02cm})\right\rangle$ is used. The matrix elements $\left \langle B^{+}(p_2) \left|\bar{u} \gamma^\mu  c\right|B_c^{+}(p_1)\right\rangle$ and $\left \langle B^{-}(p_2) \left|\bar{c} \gamma^\mu  u\right|B_c^{-}(p_1)\right\rangle$ are parameterized in terms of various form factors as~\cite{Cooper:2020wnj}
\begin{align}
\left \langle B^{+} (p_2) \left|\bar{u} \gamma^\mu  c\right|B^+_{c}(p_1)\right\rangle=f_0(q_t^2)\left(\frac{p_{1}^2-p_{2}^2}{q_t^2}{q_t^{\mu}}\right)+f_{+}(q_t^2)\left(p_1^{\mu}+p_2^{\mu}-\frac{p_{1}^2-p_{2}^2}{q_t^2}{q_t^{\mu}}\right),\label{Eq:formfactor1} \\
\left \langle B^{-} (p_2) \left|\bar{c} \gamma^\mu  u\right|B^-_{c}(p_1)\right\rangle=-f_0(q_t^2)\left(\frac{p_{1}^2-p_{2}^2}{q_t^2}{q_t^{\mu}}\right)-f_{+}(q_t^2)\left(p_1^{\mu}+p_2^{\mu}-\frac{p_{1}^2-p_{2}^2}{q_t^2}{q_t^{\mu}}\right),  \label{Eq:formfactor2}
\end{align}
where $q_t=p_1-p_2$ is the 4-momentum transfer. The form factors $f_0(q_t^2)$ and $f_+(q_t^2)$ read
\begin{align}
f_0(q_t^2)=\frac{1}{1-\frac{q_t^2}{m_{D_0^\ast}^2}}\left(b_0^{(0)}+b_0^{(1)}Z_p+b_0^{(2)}Z_p^2+b_0^{(3)}Z_p^3\right), \label{Eq:fzero}\\
f_+(q_t^2)=\frac{1}{1-\frac{q_t^2}{m_{D^\ast}^2}}\left(b_+^{(0)}+b_+^{(1)}Z_p+b_+^{(2)}Z_p^2+b_+^{(3)}Z_p^3\right), \label{Eq:fplus}
\end{align}
here, $m_{D_0^\ast}$ and $m_{D^\ast}$ is respectively the mass of the neutral scalar meson $D_0^\ast(2300)^0$ and the neutral vector  meson $D^{\ast}(2007)^0$. The variable $Z_p$ is defined as
\begin{align}
Z_p=\left(\frac{\sqrt{t_{+}-q_t^2}-\sqrt{t_+}}{\sqrt{t_{+}-q_t^2}+\sqrt{t_+}}\right)\cdot\left|\frac{\sqrt{t_{+}-m_{D^\ast}^2}+\sqrt{t_+}}{\sqrt{t_{+}-m_{D^\ast}^2}-\sqrt{t_+}}\right|, \label{Eq:zpfunction}
\end{align}
with $t_+={\left(m_{B_c^+}+m_{B^+}\right)}^2$. The polynomial coefficients $b_{0,+}^{(i)} (i=0,1,2,3)$ appearing in  Eq.(\ref{Eq:fzero}) and Eq.(\ref{Eq:fplus}) have been calculated in Ref.~\cite{Cooper:2020wnj}
\begin{align}
b_{0}^{(0)}=0.548\pm0.023,~~~~~b_{0}^{(1)}=-0.19\pm0.22,~~~~~b_{0}^{(2)}=0.05\pm0.74,~~~~~ b_{0}^{(3)}=0, \label{Eq:coefficients1}\\
b_{+}^{(0)}=0.548\pm0.023,~~~~~b_{+}^{(1)}=-0.48\pm0.21,~~~~~b_{+}^{(2)}=0.12\pm0.77,~~~~~ b_{+}^{(3)}=0.\label{Eq:coefficients2}
\end{align}

Now, we can proceed to calculate the hadronic matrix elements $\left \langle K^0 (q_t)B^{\pm}(p_2) \left|\mathcal H_{\rm eff}^{fac}\right|B^{\pm}_{c}(p_1)\right\rangle$ and $\left \langle \bar{K}^0 (q_t)B^{\pm}(p_2) \left|\mathcal H_{\rm eff}^{fac}\right|B^{\pm}_{c}(p_1)\right\rangle$. Because of the parity conservation in strong interaction, the matrix elements $\left \langle K^0 (q_t) \left|\bar{d} \gamma^\mu s\right|0\right\rangle$, $\left \langle B^{+} (p_2) \left|\bar{u} \gamma^\mu \gamma^5 c\right|B^+_{c}(p_1)\right\rangle$ and their charge conjugate have no contribution to the corresponding the hadronic matrix elements. By combining Eq.(\ref{Eq:effhamilton2}), Eq.(\ref{Eq:kdecaycons}), Eq.(\ref{Eq:kbdecaycons}), Eq.(\ref{Eq:formfactor1}), Eq.(\ref{Eq:formfactor2}) and $q_t=p_1-p_2$, we can obtain
\begin{align}
\left \langle K^0 (q_t)B^{+}(p_2) \left|\mathcal H_{\rm eff}^{fac}\right|B^{+}_{c}(p_1)\right\rangle=\frac{i{G_F} f_K}{3\sqrt{2}}V_{cd}^{\ast}V_{us}f_0(q_t^2) \left(p_{1}^2-p_{2}^2\right),\label{Eq:hadmatrixele1}\\
\left \langle \bar{K}^0 (q_t)B^{+}(p_2) \left|\mathcal H_{\rm eff}^{fac}\right|B^{+}_{c}(p_1)\right\rangle=\frac{i{G_F} f_K}{3\sqrt{2}}V_{cs}^{\ast}V_{ud}f_0(q_t^2) \left(p_{1}^2-p_{2}^2\right),\label{Eq:hadmatrixele2}\\
\left \langle K^0 (q_t)B^{-}(p_2) \left|\mathcal H_{\rm eff}^{fac}\right|B^{-}_{c}(p_1)\right\rangle=\frac{i{G_F} f_K}{3\sqrt{2}}V_{cs}V_{ud}^{\ast}f_0(q_t^2) \left(p_{1}^2-p_{2}^2\right),\label{Eq:hadmatrixele3}\\
\left \langle \bar{K}^0 (q_t)B^{-}(p_2) \left|\mathcal H_{\rm eff}^{fac}\right|B^{-}_{c}(p_1)\right\rangle=\frac{i{G_F} f_K}{3\sqrt{2}}V_{cd}V_{us}^{\ast}f_0(q_t^2) \left(p_{1}^2-p_{2}^2\right).\label{Eq:hadmatrixele4}
\end{align}

In the $K^0-\bar{K}^0$ system, the two mass eigenstates, $K^0_S$ of mass $m_S$ and width $\Gamma_S$ and $K^0_L$ of mass $m_L$ and width $\Gamma_L$, are linear combinations of the flavor eigenstates $K^0$ and $\bar{K}^0$~\cite{Zyla:2020zbs}
\begin{align}
\left | K^0_L \right\rangle=p\sqrt{1+z}\left | K^0\right\rangle-q\sqrt{1-z}\left | \bar{K}^0\right\rangle,\label{Eq:kldefinition}\\
\left | K^0_S \right\rangle=p\sqrt{1-z}\left | K^0\right\rangle+q\sqrt{1+z}\left | \bar{K}^0\right\rangle,\label{Eq:ksdefinition}
\end{align}
where $p$, $q$ and $z$ are complex mixing parameters. CPT invariant requires $z=0$, while CP conservation requires both $p=q=\sqrt{2}/2$ and $z=0$. The mass and width eigenstates, $K^0_{S,L}$, may also be described with the popular notations
\begin{align}
\left | K^0_L \right\rangle=\frac{1+\epsilon-\delta}{\sqrt{2(1+|\epsilon-\delta|^2)}}\left | K^0\right\rangle-\frac{1-\epsilon+\delta}{\sqrt{2(1+|\epsilon-\delta|^2)}}\left | \bar{K}^0\right\rangle,\label{Eq:kldefinition2}\\
\left | K^0_S \right\rangle=\frac{1+\epsilon+\delta}{\sqrt{2(1+|\epsilon+\delta|^2)}}\left | K^0\right\rangle+\frac{1-\epsilon-\delta}{\sqrt{2(1+|\epsilon+\delta|^2)}}\left | \bar{K}^0\right\rangle,\label{Eq:ksdefinition2}
\end{align}
where the complex parameter $\epsilon$ signifies deviation of the mass eigenstates from the CP eigenstates, $\delta$ is the CPT violating  complex parameter. The parameters $p$, $q$ and $z$ can be expressed in terms of $\epsilon$ and $\delta$ (neglecting terms of $\epsilon\delta$ and $\mathcal{O}(\delta)$)
\begin{align}
p=\frac{1+\epsilon}{\sqrt{2(1+|\epsilon|^2)}},~~~~~~~~~~q=\frac{1-\epsilon}{\sqrt{2(1+|\epsilon|^2)}},~~~~~~~~~~z=-2\delta.\label{Eq:ksdefinition2}
\end{align}
In this paper, we adopt the notations $p$, $q$ and $z$ in order to ensuring the rephasing invariance of all analytical results~\cite{Lipkin:1999qz}.

The time-evolved states of the $K^0-\bar{K}^0$ system can be expressed by the mass eigenstates
\begin{align}
\left | K^0_{phys} (t) \right\rangle=\frac{\sqrt{1+z}}{2p} e^{-i m_L t -\frac{1}{2}\Gamma_L t } \left | K^0_L \right\rangle +\frac{\sqrt{1-z}}{2p} e^{-i m_S t -\frac{1}{2}\Gamma_S t } \left | K^0_S \right\rangle,\label{Eq:kzphysdef}\\
\left | \bar{K}^0_{phys} (t) \right\rangle=-\frac{\sqrt{1-z}}{2q} e^{-i m_L t -\frac{1}{2}\Gamma_L t } \left | K^0_L \right\rangle +\frac{\sqrt{1+z}}{2q} e^{-i m_S t -\frac{1}{2}\Gamma_S t } \left | K^0_S \right\rangle.\label{Eq:kbphysdef}
\end{align}
With Eq.(\ref{Eq:kzphysdef}) and Eq.(\ref{Eq:kbphysdef}), the time-dependent amplitudes of the cascade decays $B^{\pm}_{c}\rightarrow B^{\pm}K^0\rightarrow B^{\pm} f_{K^0}$ can be written as
\begin{align}
&A\left(B^{\pm}_{c}\rightarrow B^{\pm}K^0(t)\rightarrow B^{\pm} f_{K^0} (t)\right)\nonumber\\
&~~~~~~~~~=\left \langle K^0 (q_t)B^{\pm}(p_2) \left|\mathcal H_{\rm eff}^{fac}\right|B^{\pm}_{c}(p_1)\right\rangle \cdot A(K^0_{phys} (t)\rightarrow f_{K^0})\nonumber\\
&~~~~~~~~~~+ \left \langle \bar{K}^0 (q_t)B^{\pm}(p_2) \left|\mathcal H_{\rm eff}^{fac}\right|B^{\pm}_{c}(p_1)\right\rangle \cdot A(\bar{K}^0_{phys} (t)\rightarrow f_{K^0}),\label{Eq:amtidedefi}
\end{align}
where $f_{K^0}$ denotes the final state from the decay of the $K^0$ or $\bar{K}^0$ meson. $A(K^0_{phys} (t)\rightarrow f_{K^0})$ and $A(\bar{K}^0_{phys} (t)\rightarrow f_{K^0})$ denotes the amplitude of the $K^0_{phys} (t)\rightarrow f_{K^0}$ and $\bar{K}^0_{phys} (t)\rightarrow f_{K^0}$ decays, respectively. They have the following forms
\begin{align}
&A(K^0_{phys} (t)\rightarrow f_{K^0})=\frac{\sqrt{1+z}}{2p} e^{-i m_L t -\frac{1}{2}\Gamma_L t }  A(K_L^0 \rightarrow f_{K^0} )+\frac{\sqrt{1-z}}{2p} e^{-i m_S t -\frac{1}{2}\Gamma_S t } A(K_S^0 \rightarrow f_{K^0} ),\label{Eq:amtikphystofkt1}\\
&A(\bar{K}^0_{phys} (t)\rightarrow f_{K^0})=-\frac{\sqrt{1-z}}{2q} e^{-i m_L t -\frac{1}{2}\Gamma_L t }  A(K_L^0 \rightarrow f_{K^0} )+\frac{\sqrt{1+z}}{2q} e^{-i m_S t -\frac{1}{2}\Gamma_S t } A(K_S^0 \rightarrow f_{K^0} ).\label{Eq:amtikphystofkt2}
\end{align}
Making use of Eqs.(\ref{Eq:hadmatrixele1}), (\ref{Eq:hadmatrixele2}), (\ref{Eq:amtidedefi}), (\ref{Eq:amtikphystofkt1}) and (\ref{Eq:amtikphystofkt2}) and performing integration over phase space, we can obtain
\begin{align}
&\Gamma(B^{+}_{c}\rightarrow B^{+}K(t)\rightarrow B^{+} f_{K^0} (t))=\frac{G_F^2 f_K^2 {|f_0(m_{K^0}^2)|}^2}{1152\pi |p|^2 |q|^2 m_{B_c^+}^3 }\cdot f(m_{B_c^+},m_{B^+} ,m_{K^0})\cdot (m_{B_c^+}^2-m_{B^+}^2)^2\nonumber\\
&~~~~~~~~~~~~~~~~~~~~~~~~~~~~~~~~~~~~~~~~~\cdot\left[ e^{-\Gamma_L t}\cdot \Gamma(K_L^0\rightarrow f_{K^0})\cdot g_{K_L^0}+e^{-\Gamma_S t}\cdot \Gamma(K_S^0\rightarrow f_{K^0})\cdot g_{K_S^0}\right.\nonumber\\
&~~~~~~~~~~~~~~~~~~~~~~~~~~~~~~~~~~~~~~~~~~~~\left.+e^{-i \Delta m t-\Gamma t}\cdot A^{*}(K_S^0\rightarrow f_{K^0})  \cdot A(K_L^0\rightarrow f_{K^0})\cdot g_{K_S^0-K_L^0}\right.\nonumber\\
&~~~~~~~~~~~~~~~~~~~~~~~~~~~~~~~~~~~~~~~~~~~~\left.+e^{i \Delta m t-\Gamma t}\cdot A(K_S^0\rightarrow f_{K^0})  \cdot A^{*}(K_L^0\rightarrow f_{K^0})\cdot g_{K_S^0-K_L^0}^*\right],
\label{Eq:decaywidthbct1}
\end{align}
where $m_r$ with $r=B_c^+, B^+, K^0$ denotes the mass of the resonance, $\Delta m$ denotes the difference in masses of $K_L^0$ and $K_S^0$, $\Gamma$ denotes the average in widths of $K_L^0$ and $K_S^0$
\begin{align}
\Delta m=m_L-m_S,~~~~~~~~~~~~~~~~~~~~\Gamma=\frac{\Gamma_L +\Gamma_S}{2}.
\label{Eq:dimassavegamma}
\end{align}
The terms $g_{K_L^0}$, $g_{K_S^0}$ and $g_{K_S^0-K_L^0}$ are related to the effect of the $K_L^0$ decay, the $K_S^0$ decay and their interference, respectively, and have the following forms
\begin{align}
&g_{K_L^0}= \left[|q|^2 |\sqrt{1+z}|^2 |V_{cd}|^2 |V_{us}|^2+|p|^2 |\sqrt{1-z}|^2 |V_{cs}|^2 |V_{ud}|^2\right.\nonumber\\
&~~~~~~~~~~~~~~~~~~~~\left. -p q^{*} \sqrt{1-z}(\sqrt{1+z})^{*} V_{cs}^{*} V_{ud} V_{cd}V_{us}^{*} -p^{*} q \sqrt{1+z} (\sqrt{1-z})^{*}  V_{cd}^{*} V_{us} V_{cs} V_{ud}^{*}\right],\label{Eq:dewigkldef}\\
&g_{K_S^0}= \left[|q|^2 |\sqrt{1-z}|^2 |V_{cd}|^2 |V_{us}|^2+|p|^2 |\sqrt{1+z}|^2 |V_{cs}|^2 |V_{ud}|^2\right.\nonumber\\
&~~~~~~~~~~~~~~~~~~~~\left. +p q^{*} \sqrt{1+z}(\sqrt{1-z})^{*} V_{cs}^{*} V_{ud} V_{cd}V_{us}^{*}+p^{*} q \sqrt{1-z} (\sqrt{1+z})^{*}  V_{cd}^{*} V_{us} V_{cs} V_{ud}^{*} \right],\label{Eq:dewigksdef}\\
&g_{K_S^0-K_L^0}=\left[|q|^2 \sqrt{1+z} (\sqrt{1-z})^{*}|V_{cd}|^2 |V_{us}|^2-|p|^2 \sqrt{1-z} (\sqrt{1+z})^{*}|V_{cs}|^2 |V_{ud}|^2\right.\nonumber\\
&~~~~~~~~~~~~~~~~~~~~\left. -p q^{*}|\sqrt{1-z}|^2 V_{cs}^{*} V_{ud} V_{cd}V_{us}^{*}+ p^{*} q |\sqrt{1+z}|^2V_{cd}^{*} V_{us} V_{cs} V_{ud}^{*} \right],\label{Eq:dewigksldef}
\end{align}
The function $f(x,y,z)$ in Eq.(\ref{Eq:decaywidthbct1}) is defined as
\begin{align}
f(x,y,z)=\sqrt{x^4+y^4+z^4-2x^2 y^2-2x^2 z^2-2y^2 z^2}.
\label{Eq:phasespacefunc}
\end{align}
In experiment, the $K_S^0$ state is defined via a final state $\pi^+\pi^-$  with $m_{\pi\pi} \approx m_S$ and a time difference between the $B_c$ decay and the $K_S^0$ decay~\cite{Grossman:2011zk}. By taking into account these experimental features, the partial decay width for the $B_c^+\rightarrow B^{+}K_S^0$ decay can be defined as
\begin{align}
\Gamma(B_c^+\rightarrow B^{+}&K_S^0)=\frac{\int_{t_0}^{t_1} \Gamma(B^{+}_{c}\rightarrow B^{+}K(t)\rightarrow B^{+} \pi^+\pi^- (t)) dt}{\left(e^{-\Gamma_S t_0}-e^{-\Gamma_S t_1}\right)\cdot{\mathcal B}(K^0_S\rightarrow \pi^+\pi^- ) },\label{Eq:bcpbpkszgadef}
\end{align}
where $t_0=0.1\tau_S$ and $t_1=2\tau_S\sim 20 \tau_S$ with $\tau_S$ is the $K_S^0$ lifetime, we adopt $t_1=10 \tau_S$ in our calculation. Combining Eq.(\ref{Eq:decaywidthbct1}) and  Eq.(\ref{Eq:bcpbpkszgadef}), we can obtain
\begin{align}
&\Gamma(B_c^+\rightarrow B^{+}K_S^0)=\frac{G_F^2 f_K^2 {|f_0(m_{S}^2)|}^2}{1152\pi |p|^2 |q|^2 m_{B_c^+}^3 }\cdot f(m_{B_c^+},m_{B^+} ,m_{S})\cdot (m_{B_c^+}^2-m_{B^+}^2)^2\nonumber\\
&~~~~~~~~~~~~~~~~~~~~\cdot\left[ g_{K_S^0}+\frac{e^{-\Gamma_L t_0}-e^{-\Gamma_L t_1}}{e^{-\Gamma_S t_0}-e^{-\Gamma_S t_1}} \cdot \frac{{\mathcal B}(K^0_L\rightarrow \pi^+\pi^- )}{{\mathcal B}(K^0_S\rightarrow \pi^+\pi^- )}\cdot g_{K_L^0}\right.\nonumber\\
&~~~~~~~~~~~~~~~~~~~~~~\left.+2Re\left(\frac{e^{-i \Delta m t_0-\Gamma t_0}-e^{-i \Delta m t_1-\Gamma t_1}}{e^{-\Gamma_S t_0}-e^{-\Gamma_S t_1}}\cdot \frac{\Gamma_S}{\Gamma+i \Delta m }\cdot \frac{A(K_L^0\rightarrow \pi^+\pi^-)}{A(K_S^0\rightarrow \pi^+\pi^-)}\cdot g_{K_S^0-K_L^0}\right)\right].\label{Eq:bcpbpksdeft1}
\end{align}
From Particle Data Group~\cite{Zyla:2020zbs}, $\Gamma_L/\Gamma_S =(1.75\pm0.01)\times 10^{-3}$. If we adopt $t_1=10 \tau_S=10/\Gamma_S$, then we can obtain
\begin{align}
&\frac{e^{-\Gamma_L t_0}-e^{-\Gamma_L t_1}}{e^{-\Gamma_S t_0}-e^{-\Gamma_S t_1}} \approx 0.019, \label{Eq:valtimekskl}
\end{align}
the value of ${\mathcal B}(K^0_L\rightarrow \pi^+\pi^- )/{\mathcal B}(K^0_S\rightarrow \pi^+\pi^- )$ can also obtained from Particle Data Group~\cite{Zyla:2020zbs}
\begin{align}
&\frac{{\mathcal B}(K^0_L\rightarrow \pi^+\pi^- )}{{\mathcal B}(K^0_S\rightarrow \pi^+\pi^- )}= (2.84\pm 0.01)\times 10^{-3}, \label{Eq:valraksltopipi}
\end{align}
so the second term in the bracket in Eq.(\ref{Eq:bcpbpksdeft1}), which corresponds to the effect of the $K_L^0$ decay, can be neglected. Using Eq.(\ref{Eq:kldefinition}), Eq.(\ref{Eq:ksdefinition}) and neglecting the tiny direct CP asymmetry in the $K^0\rightarrow \pi^+\pi^-$ decay, we can derive
\begin{align}
&\frac{A(K_L^0\rightarrow \pi^+\pi^-)}{A(K_S^0\rightarrow \pi^+\pi^-)}= \frac{p\sqrt{1+z}-q \sqrt{1-z}}{p\sqrt{1-z}+q \sqrt{1+z}}. \label{Eq:ratiokslpipiam}
\end{align}
For convenience, we introduce the following substitution
\begin{align}
&t_{K^0_S-K_L^0}=\frac{e^{-i \Delta m t_0-\Gamma t_0}-e^{-i \Delta m t_1-\Gamma t_1}}{e^{-\Gamma_S t_0}-e^{-\Gamma_S t_1}}\cdot \frac{\Gamma_S}{\Gamma+i \Delta m }.\label{Eq:substitigamma}
\end{align}
Combining Eqs.(\ref{Eq:dewigksdef}), (\ref{Eq:dewigksldef}), (\ref{Eq:bcpbpksdeft1}), (\ref{Eq:valtimekskl}), (\ref{Eq:valraksltopipi}) and (\ref{Eq:ratiokslpipiam}), we can derive
\begin{align}
\Gamma(B_c^+\rightarrow B^{+}&K_S^0)=\frac{G_F^2 f_K^2 {|f_0(m_{S}^2)|}^2}{1152\pi |p|^2 |q|^2 m_{B_c^+}^3 }\cdot f(m_{B_c^+},m_{B^+} ,m_{S})\cdot (m_{B_c^+}^2-m_{B^+}^2)^2\nonumber\\
&\cdot\bigg\{ \left[|q|^2 |\sqrt{1-z}|^2 |V_{cd}|^2 |V_{us}|^2+|p|^2 |\sqrt{1+z}|^2 |V_{cs}|^2 |V_{ud}|^2\right.\nonumber\\
&\left. +p q^{*} \sqrt{1+z}(\sqrt{1-z})^{*} V_{cs}^{*} V_{ud} V_{cd}V_{us}^{*} +p^{*} q \sqrt{1-z} (\sqrt{1+z})^{*}  V_{cd}^{*} V_{us} V_{cs} V_{ud}^{*}\right]\nonumber\\
&+2Re\left(t_{K^0_S-K_L^0}\cdot\frac{p\sqrt{1+z}-q \sqrt{1-z}}{p\sqrt{1-z}+q \sqrt{1+z}}\cdot \left[|q|^2 \sqrt{1+z} (\sqrt{1-z})^{*}|V_{cd}|^2 |V_{us}|^2\right.\right.\nonumber\\
&-|p|^2 \sqrt{1-z} (\sqrt{1+z})^{*}|V_{cs}|^2 |V_{ud}|^2-p q^{*}|\sqrt{1-z}|^2 V_{cs}^{*} V_{ud} V_{cd}V_{us}^{*}\nonumber\\
&\left.  \left.+ p^{*} q |\sqrt{1+z}|^2V_{cd}^{*} V_{us} V_{cs} V_{ud}^{*} \right]\bigg)\right\}.\label{Eq:decaywidth1}
\end{align}
Similarly, we can obtain the partial decay width for the $B_c^-\rightarrow B^{-}K_S^0$ decay
\begin{align}
\Gamma(B_c^-\rightarrow B^{-}&K_S^0)=\frac{G_F^2 f_K^2 {|f_0(m_S^2)|}^2}{1152\pi |p|^2 |q|^2 m_{B_c^+}^3}\cdot f(m_{B_c^+},m_{B^+} ,m_S)\cdot (m_{B_c^+}^2-m_{B^+}^2)^2\nonumber\\
&\cdot\bigg\{\left[|q|^2 |\sqrt{1-z}|^2 |V_{cs}|^2 |V_{ud}|^2+|p|^2 |\sqrt{1+z}|^2 |V_{cd}|^2 |V_{us}|^2\right.\nonumber\\
&\left.+p q^{*} \sqrt{1+z} (\sqrt{1-z})^{*}  V_{cs}^{*} V_{ud} V_{cd} V_{us}^{*}+p^{*} q \sqrt{1-z} (\sqrt{1+z})^{*} V_{cs} V_{ud}^{*} V_{cd}^{*} V_{us}\right]\nonumber\\
&+2Re\left(t_{K^0_S-K_L^0}\cdot\frac{p\sqrt{1+z}-q \sqrt{1-z}}{p\sqrt{1-z}+q \sqrt{1+z}}\cdot \left[|q|^2 \sqrt{1+z} (\sqrt{1-z})^{*}|V_{cs}|^2 |V_{ud}|^2\right.\right.\nonumber\\
&-|p|^2 \sqrt{1-z} (\sqrt{1+z})^{*}|V_{cd}|^2 |V_{us}|^2-p q^{*}|\sqrt{1-z}|^2 V_{cs}^{*} V_{ud} V_{cd}V_{us}^{*}\nonumber\\
&\left.  \left.+ p^{*} q |\sqrt{1+z}|^2V_{cd}^{*} V_{us} V_{cs} V_{ud}^{*} \right]\bigg)\right\}.\label{Eq:decaywidth2}
\end{align}
In experiment, the $K_L^0$ state is defined via a large time difference between the $B_c$ decay and the $K_L^0$ decay and mostly decay outside the detector~\cite{Cerri:2018ypt}. Basing on these experimental features, the partial decay width for the $B_c^+\rightarrow B^{+}K_L^0$ decay can be defined as
\begin{align}
\Gamma(B_c^+\rightarrow B^{+}&K_L^0)=\frac{\int_{t_2}^{+\infty} \Gamma(B^{+}_{c}\rightarrow B^{+}K(t)\rightarrow B^{+} f_{K_L^0} (t)) dt}{e^{-\Gamma_L t_2}\cdot{\mathcal B}(K^0_L\rightarrow f_{K_L^0} ) },\label{Eq:bcpbpklzgadef}
\end{align}
where $f_{K_L^0}$ denotes the final state of the $K_L^0$ decay, $t_2\geq 100\tau_S$. Combining Eq.(\ref{Eq:decaywidthbct1}) and  Eq.(\ref{Eq:bcpbpklzgadef}), we can obtain
\begin{align}
\Gamma(B_c^+\rightarrow B^{+}&K_L^0)=\frac{G_F^2 f_K^2 {|f_0(m_{L}^2)|}^2}{1152\pi |p|^2 |q|^2 m_{B_c^+}^3 }\cdot f(m_{B_c^+},m_{B^+} ,m_{L})\cdot (m_{B_c^+}^2-m_{B^+}^2)^2\nonumber\\
&~~~~~~~~~~\cdot\left[ g_{K_L^0}+e^{-(\Gamma_S-\Gamma_L) t_2} \cdot \frac{{\mathcal B}(K^0_S\rightarrow f_{K_L^0} )}{{\mathcal B}(K^0_L\rightarrow f_{K_L^0} )}\cdot g_{K_S^0}\right.\nonumber\\
&~~~~~~~~~~~~\left.+2Re\left(e^{i \Delta m t_2-\frac{\Gamma_S-\Gamma_L}{2} t_2}\cdot \frac{\Gamma_L}{\Gamma-i \Delta m }\cdot \frac{A(K_S^0\rightarrow f_{K_L^0})}{A(K_L^0\rightarrow f_{K_L^0})}\cdot g_{K_S^0-K_L^0}^*\right)\right].\label{Eq:bcpbpkldeft1}
\end{align}
Using the result from Particle Data Group~\cite{Zyla:2020zbs}: $\Gamma_L/\Gamma_S =(1.75\pm0.01)\times 10^{-3}$, we can obtain
\begin{align}
&e^{-(\Gamma_S-\Gamma_L) t_2}\le 4.4\times 10^{-44},
~~~~~~~~~~~~~~~~~~~~~~~~~~e^{-\frac{\Gamma_S-\Gamma_L}{2} t_2}\le 2.1\times 10^{-22},\label{Eq:valuettexpimp}
\end{align}
so the second and the third terms in the bracket in Eq.(\ref{Eq:bcpbpkldeft1}), which corresponds to the effects of the $K_S^0$ decay and the interference between the $K_S^0$ decay and the the $K_L^0$ decay, can be neglected. Substituting Eq.(\ref{Eq:dewigkldef}) into Eq.(\ref{Eq:bcpbpkldeft1}) and neglecting the second and the third terms in the bracket in Eq.(\ref{Eq:bcpbpkldeft1}), we obtain
\begin{align}
\Gamma(B_c^+\rightarrow B^{+}&K_L^0)=\frac{G_F^2 f_K^2 {|f_0(m_{L}^2)|}^2}{1152\pi |p|^2 |q|^2 m_{B_c^+}^3 }\cdot f(m_{B_c^+},m_{B^+} ,m_{L})\cdot (m_{B_c^+}^2-m_{B^+}^2)^2\nonumber\\
&~\cdot\left[|q|^2 |\sqrt{1+z}|^2 |V_{cd}|^2 |V_{us}|^2+|p|^2 |\sqrt{1-z}|^2 |V_{cs}|^2 |V_{ud}|^2\right.\nonumber\\
&~\left. -p q^{*} \sqrt{1-z}(\sqrt{1+z})^{*} V_{cs}^{*} V_{ud} V_{cd}V_{us}^{*} -p^{*} q \sqrt{1+z} (\sqrt{1-z})^{*}  V_{cd}^{*} V_{us} V_{cs} V_{ud}^{*}\right].\label{Eq:decaywidth3}
\end{align}
Similarly, we can obtain the partial decay width for the $B_c^-\rightarrow B^{-}K_L^0$ decay
\begin{align}
\Gamma(B_c^-\rightarrow B^{-}&K_L^0)=\frac{G_F^2 f_K^2 {|f_0(m_L^2)|}^2}{1152\pi |p|^2 |q|^2 m_{B_c^+}^3}\cdot f(m_{B_c^+},m_{B^+} ,m_L)\cdot (m_{B_c^+}^2-m_{B^+}^2)^2\nonumber\\
&\cdot\left[|q|^2 |\sqrt{1+z}|^2 |V_{cs}|^2 |V_{ud}|^2+|p|^2 |\sqrt{1-z}|^2 |V_{cd}|^2 |V_{us}|^2\right.\nonumber\\
&\left.-p q^{*} \sqrt{1-z}  (\sqrt{1+z})^{*}V_{cs}^{*} V_{ud} V_{cd} V_{us}^{*}-p^{*} q  \sqrt{1+z} (\sqrt{1-z})^{*} V_{cs} V_{ud}^{*} V_{cd}^{*} V_{us}\right].\label{Eq:decaywidth4}
\end{align}
From Eqs.(\ref{Eq:decaywidth1}), (\ref{Eq:decaywidth2}), (\ref{Eq:decaywidth3}) and (\ref{Eq:decaywidth4}), we can obtain the branching ratios of $B_c^\pm\rightarrow B^{\pm}K_{S,L}^0$ decays, which are the products of the partial decay widths for these decays and the mean life of $B_c$ meson.
\section{CP violations, $K_{S}^{0}$-$K_{L}^{0}$  asymmetries and CPT violations }
\label{sec:cpvcptvkskl}
In the $B_c^\pm\rightarrow B^{\pm}K_{S,L}^0\rightarrow f_{B^\pm} f_{K_{S,L}^0}$ decays, the time-independent CP asymmetries are defined as
\begin{align}
{\mathcal A}_{CP}^{K_{S,L}^{0}}=\frac{\Gamma(B_c^-\rightarrow B^{-}K_{S,L}^0)\cdot\Gamma(B^{-}\rightarrow f_{B^-})-\Gamma(B_c^+\rightarrow B^{+}K_{S,L}^0)\cdot\Gamma(B^{+}\rightarrow \bar{f}_{B^-})}{\Gamma(B_c^-\rightarrow B^{-}K_{S,L}^0)\cdot\Gamma(B^{-}\rightarrow f_{B^-})+\Gamma(B_c^+\rightarrow B^{+}K_{S,L}^0)\cdot\Gamma(B^{+}\rightarrow \bar{f}_{B^-})},\label{Eq:cpasymmksl}
\end{align}
here, $f_{r}$ with  $r=B^\pm, K_{S,L}^0$ denote the final states from the decays of the $B^\pm$ and $K_{S,L}^0$ mesons, $\bar{f}_{B^\pm}$ are the CP-conjugate states of $f_{B^\pm}$, $\Gamma(B^{-}\rightarrow f_{B^-})$ and $\Gamma(B^{+}\rightarrow \bar{f}_{B^-})$ is the partial decay width of the $B^{-}\rightarrow f_{B^-}$ and $B^{+}\rightarrow \bar{f}_{B^-}$ decays respectively. Using the relations in Eqs.(\ref{Eq:bianhuanrelation1}) and (\ref{Eq:bianhuanrelation2}) listed in Appendix~\ref{sec:appendixcpcal}, we can obtain
\begin{align}
{\mathcal A}_{CP}^{K_{S,L}^{0}}=\frac{\Gamma(B^{-}\rightarrow f_{B^-})-\Gamma(B^{+}\rightarrow \bar{f}_{B^-})}{\Gamma(B^{-}\rightarrow f_{B^-})+\Gamma(B^{+}\rightarrow \bar{f}_{B^-})}+\frac{\Gamma(B_c^-\rightarrow B^{-}K_{S,L}^0)-\Gamma(B_c^+\rightarrow B^{+}K_{S,L}^0)}{\Gamma(B_c^-\rightarrow B^{-}K_{S,L}^0)+\Gamma(B_c^+\rightarrow B^{+}K_{S,L}^0)},\label{Eq:cpasymmsimksl}
\end{align}
where $\left(\Gamma(B_c^-\rightarrow B^{-}K_{S,L}^0)-\Gamma(B_c^+\rightarrow B^{+}K_{S,L}^0)\right)/\left(\Gamma(B_c^-\rightarrow B^{-}K_{S,L}^0)+\Gamma(B_c^+\rightarrow B^{+}K_{S,L}^0)\right)$ are the CP asymmetries in the $B_c^\pm\rightarrow B^{\pm}K_{S,L}^0$ decays and hereinafter denoted as ${\mathcal A}_{CP}\left(B_c^\pm\rightarrow B^{\pm}K_{S,L}^0\right)$. Under the assumption of CPT invariance, ${\mathcal A}_{CP}\left(B_c^\pm\rightarrow B^{\pm}K_{S,L}^0\right)$ can be calculated from Eqs.(\ref{Eq:decaywidth1}), (\ref{Eq:decaywidth2}), (\ref{Eq:decaywidth3}) and (\ref{Eq:decaywidth4})
\begin{align}
{\mathcal A}_{CP}&\left(B_c^\pm\rightarrow  B^{\pm}K_{S}^0\right)\nonumber\\
&=\frac{\left({\left|q\right|}^2-{\left|p\right|}^2\right)\cdot\left({\left|V_{cs}\right|}^2{\left|V_{ud}\right|}^2-{\left|V_{cd}\right|}^2{\left|V_{us}\right|}^2\right) }{\left({\left|p\right|}^2+{\left|q\right|}^2\right)\cdot\left({\left|V_{cs}\right|}^2{\left|V_{ud}\right|}^2+{\left|V_{cd}\right|}^2{\left|V_{us}\right|}^2\right)+ 2pq^{*}V_{cs}^{*} V_{ud} V_{cd} V_{us}^{*}+ 2p^{*}qV_{cs} V_{ud}^{*} V_{cd}^{*} V_{us} }\nonumber\\
&+\frac{2\left({\left|V_{cs}\right|}^2{\left|V_{ud}\right|}^2-{\left|V_{cd}\right|}^2{\left|V_{us}\right|}^2\right)\cdot \left({\left|p\right|}^2+{\left|q\right|}^2\right) \cdot Re\left(t_{K^0_S-K_L^0}\cdot\frac{p-q}{p+q}\right) }{\left({\left|p\right|}^2+{\left|q\right|}^2\right)\cdot\left({\left|V_{cs}\right|}^2{\left|V_{ud}\right|}^2+{\left|V_{cd}\right|}^2{\left|V_{us}\right|}^2\right)+ 2pq^{*}V_{cs}^{*} V_{ud} V_{cd} V_{us}^{*}+ 2p^{*}qV_{cs} V_{ud}^{*} V_{cd}^{*} V_{us} },\label{Eq:cpbcpmtobpmks}
\end{align}
\begin{align}
{\mathcal A}_{CP}&\left(B_c^\pm\rightarrow  B^{\pm}K_{L}^0\right)\nonumber\\
&=\frac{\left({\left|q\right|}^2-{\left|p\right|}^2\right)\cdot\left({\left|V_{cs}\right|}^2{\left|V_{ud}\right|}^2-{\left|V_{cd}\right|}^2{\left|V_{us}\right|}^2\right) }{\left({\left|p\right|}^2+{\left|q\right|}^2\right)\cdot\left({\left|V_{cs}\right|}^2{\left|V_{ud}\right|}^2+{\left|V_{cd}\right|}^2{\left|V_{us}\right|}^2\right)- 2pq^{*}V_{cs}^{*} V_{ud} V_{cd} V_{us}^{*}- 2p^{*}qV_{cs} V_{ud}^{*} V_{cd}^{*} V_{us} }.\label{Eq:cpbcpmtobpmkl}
\end{align}
From Eq.(\ref{Eq:cpbcpmtobpmks}) and Eq.(\ref{Eq:cpbcpmtobpmkl}), we can see that the dominant contribution to the CP asymmetries in the $B_{c}^{\pm}\rightarrow B^{\pm} K_{S,L}^{0}$ decays arise indirectly from $K^0-{\bar{K}}^0$ mixing, the reason is that the strong phase difference between $B_c^\pm\rightarrow  B^{\pm}K^0$  and $B_c^\pm\rightarrow  B^{\pm}\bar{K}^0$ magnitudes is absent at tree level in the SM.

In Eq.(\ref{Eq:cpasymmsimksl}), $\left(\Gamma(B^{-}\rightarrow f_{B^-})-\Gamma(B^{+}\rightarrow \bar{f}_{B^-})\right)/\left(\Gamma(B^{-}\rightarrow f_{B^-})+\Gamma(B^{+}\rightarrow \bar{f}_{B^-})\right)$ is the CP asymmetry in $B^-\rightarrow f_{B^-}$ and $B^+\rightarrow \bar{f}_{B^-}$ decays and hereinafter denoted as ${\mathcal A}_{CP}\left(B^-\rightarrow f_{B^-}\right)$. For the semileptonic decays of the $B^{\pm}$ to charmed meson, such as $B^{\pm}\rightarrow \bar{D}^0 l^{\pm}\nu_l$ and $B^{\pm}\rightarrow \bar{D}^{\ast}(2007)^0 l^{\pm}\nu_l$, the direct CP violations don't occur in the SM~\cite{D0:2016xvr}, so ${\mathcal A}_{CP}^{K_{S}^{0}}$ and ${\mathcal A}_{CP}^{K_{L}^{0}}$ are only determined by ${\mathcal A}_{CP}\left(B_c^\pm\rightarrow  B^{\pm}K_{S,L}^0\right)$. Here, we note that ${\mathcal A}_{CP}\left(B^-\rightarrow f_{B^-}\right)$ should also receive the contribution from the CP violations in charm meson decays, for example, the CP violations in $B^{\pm}\rightarrow \bar{D}^0 l^{\pm}\nu_l\rightarrow \bar{f}_{D} l^{\pm}\nu_l$ decay chains can be derived as
\begin{align}
{\mathcal A}_{CP}&\left(B^{-}\rightarrow  D^0 l^- \bar{\nu}_l\right)\nonumber\\
&=\frac{\Gamma(B^{-}\rightarrow  D^0 l^- \bar{\nu}_l)-\Gamma(B^{+}\rightarrow  \bar{D}^0 l^+ \nu_l)}{\Gamma(B^{-}\rightarrow  D^0 l^- \bar{\nu}_l)+\Gamma(B^{+}\rightarrow  \bar{D}^0 l^+ \nu_l)}+\frac{\Gamma(D^0\rightarrow f_{D})-\Gamma(\bar{D}^0\rightarrow \bar{f}_{D})}{\Gamma(D^0\rightarrow f_{D})+\Gamma(\bar{D}^0\rightarrow \bar{f}_{D})},\label{Eq:cpbplusdzlvl}
\end{align}
here, $f_{D}$ and $\bar{f}_{D}$ denote the final states of $D^0$ and $\bar{D}^0$ decays, respectively. Fortunately, the second term on the right side of Eq.(\ref{Eq:cpbplusdzlvl}) is zero in the semileptonic and leptonic decays of the D mesons in the SM. However, because the final states of $B^{\pm}\rightarrow  D$ semileptonic decays followed by the semileptonic and leptonic charm decays contain two neutrinos, the experiment environment is imperfect. The CP asymmetries in the nonleptonic decays of the $B^{\pm}$ mesons have nonzero values, some of these values can reach the order of $10^{-1}$, so the CP asymmetries in the nonleptonic decays of the $B^{\pm}$ mesons couldn't be neglected, we will discuss the possibility to observe the CP violations in the $B_c^{\pm}\rightarrow B^{\pm}K_{S,L}^0$ decays followed by the nonleptonic decays of the $B^{\pm}$ mesons in section~\ref{sec:numcalution}.

Next, we turn to calculate the $K_{S}^0-K_{L}^0$ asymmetries in the $B_c^{+}\rightarrow B^{+}K_{S,L}^0$ and $B_c^{-}\rightarrow B^{-}K_{S,L}^0$ decays, which are defined as~\cite{Yu:2017oky,Bigi:1994aw,Wang:2017ksn,CLEO:2007rhw}
\begin{align}
R\left(B_c^{+}\rightarrow  B^+ K_{S,L}^0 \right)=\frac{\Gamma\left(B_c^{+}\rightarrow  B^+ K_{S}^0 \right)-\Gamma\left(B_c^{+}\rightarrow  B^+ K_{L}^0 \right)}{\Gamma\left(B_c^{+}\rightarrow  B^+ K_{S}^0 \right)+\Gamma\left(B_c^{+}\rightarrow  B^+ K_{L}^0 \right)},\label{Eq:ksklasymmetry1}\\
R\left(B_c^{-}\rightarrow  B^- K_{S,L}^0 \right)=\frac{\Gamma\left(B_c^{-}\rightarrow  B^- K_{S}^0 \right)-\Gamma\left(B_c^{-}\rightarrow  B^- K_{L}^0 \right)}{\Gamma\left(B_c^{-}\rightarrow  B^- K_{S}^0 \right)+\Gamma\left(B_c^{-}\rightarrow  B^- K_{L}^0 \right)},\label{Eq:ksklasymmetry2}
\end{align}
By substituting Eqs.(\ref{Eq:decaywidth1}), (\ref{Eq:decaywidth2}), (\ref{Eq:decaywidth3}) and (\ref{Eq:decaywidth4}) into Eqs.(\ref{Eq:ksklasymmetry1})-(\ref{Eq:ksklasymmetry2}) and neglecting the $K_S^0-K_L^0$ mass difference and the CPT violation parameter $z$, we can obtain
\begin{align}
&R\left(B_c^{+}\rightarrow  B^+ K_{S,L}^0 \right)=\frac{p q^* V_{cs}^* V_{ud} V_{cd} V_{us}^* + p^* q V_{cs} V_{ud}^* V_{cd}^* V_{us}}{|q|^2|V_{cd}|^2 |V_{us}|^2 + |p|^2|V_{cs}|^2 |V_{ud}|^2}\nonumber\\
&+\frac{Re\left(t_{K^0_S-K_L^0}\cdot\frac{p-q}{p+q}\cdot\left[|q|^2 |V_{cd}|^2 |V_{us}|^2-|p|^2|V_{cs}|^2 |V_{ud}|^2
-p q^{*} V_{cs}^{*} V_{ud} V_{cd}V_{us}^{*}+ p^{*} q V_{cd}^{*} V_{us} V_{cs} V_{ud}^{*} \right] \right)}{|q|^2|V_{cd}|^2 |V_{us}|^2 + |p|^2|V_{cs}|^2 |V_{ud}|^2},\label{Eq:ksklasyresult1}\\
&R\left(B_c^{-}\rightarrow  B^- K_{S,L}^0 \right)=\frac{p q^* V_{cs}^* V_{ud} V_{cd} V_{us}^* + p^* q V_{cs} V_{ud}^* V_{cd}^* V_{us}}{|q|^2|V_{cs}|^2 |V_{ud}|^2 + |p|^2|V_{cd}|^2 |V_{us}|^2}\nonumber\\
&+\frac{Re\left(t_{K^0_S-K_L^0}\cdot\frac{p-q}{p+q}\cdot\left[|q|^2 |V_{cs}|^2 |V_{ud}|^2-|p|^2|V_{cd}|^2 |V_{us}|^2
-p q^{*} V_{cs}^{*} V_{ud} V_{cd}V_{us}^{*}+ p^{*} q V_{cd}^{*} V_{us} V_{cs} V_{ud}^{*} \right] \right)}{|q|^2|V_{cs}|^2 |V_{ud}|^2 + |p|^2|V_{cd}|^2 |V_{us}|^2},\label{Eq:ksklasyresult2}
\end{align}
here, we note that the $K_S^0-K_L^0$ mass difference is involved in the following expression in $R\left(B_c^{\pm}\rightarrow  B^{\pm} K_{S,L}^0 \right)$
\begin{align}
\frac{{|f_0(m_S^2)|}^2f(m_{B_c^+},m_{B^+},m_S)-{|f_0(m_L^2)|}^2f(m_{B_c^+},m_{B^+} ,m_L)}{{|f_0(m_S^2)|}^2 f(m_{B_c^+},m_{B^+},m_S)},\label{Eq:termofkslmdif}
\end{align}
the experimental value of the $K_S^0-K_L^0$ mass difference is $(3.48\pm0.01)\times 10^{-15} \text{GeV}$~\cite{Zyla:2020zbs}, so the above expression is of the order of $10^{-16}$ and can be safely neglected.

Now, we proceed to study the CPT violation parameter $z$ in the $B_c^{\pm}\rightarrow B^{\pm}K_{S,L}^0$ decays. From Eqs.(\ref{Eq:decaywidth1}), (\ref{Eq:decaywidth2}), (\ref{Eq:decaywidth3}) and (\ref{Eq:decaywidth4}), we can obtain
\begin{align}
&\Gamma\left(B_c^{-}\rightarrow  B^- K_{S}^0 \right)-\Gamma\left(B_c^{+}\rightarrow  B^+ K_{S}^0 \right)=\frac{G_F^2 f_K^2 {|f_0(m_S^2)|}^2}{1152\pi |p|^2 |q|^2 m_{B_c^+}^3}\cdot f(m_{B_c^+},m_{B^+} ,m_S)\cdot (m_{B_c^+}^2-m_{B^+}^2)^2 \nonumber\\
&~~~~~~~~\cdot\left(|V_{cs}|^2 |V_{ud}|^2 -|V_{cd}|^2 |V_{us}|^2\right)\cdot\left\{\left(|q|^2 |\sqrt{1-z}|^2 -|p|^2 |\sqrt{1+z}|^2 \right) \right.\nonumber\\
&~\left.+2Re\left(t_{K^0_S-K_L^0}\cdot\frac{p\sqrt{1+z}-q \sqrt{1-z}}{p\sqrt{1-z}+q \sqrt{1+z}}\cdot\left[|q|^2(\sqrt{1-z})^{*} \sqrt{1+z}+|p|^2\sqrt{1-z}(\sqrt{1+z})^{*}  \right]\right)\right\},\label{Eq:cptnumerator1}
\end{align}
\begin{align}
\Gamma\left(B_c^{-}\rightarrow  B^- K_{L}^0 \right)-&\Gamma\left(B_c^{+}\rightarrow  B^+ K_{L}^0 \right)=\frac{G_F^2 f_K^2 {|f_0(m_L^2)|}^2}{1152\pi |p|^2 |q|^2 m_{B_c^+}^3}\cdot f(m_{B_c^+},m_{B^+} ,m_L)\cdot (m_{B_c^+}^2-m_{B^+}^2)^2 \nonumber\\
&~~~~~~~~\cdot(|q|^2 |\sqrt{1+z}|^2 -|p|^2 |\sqrt{1-z}|^2 )\cdot (|V_{cs}|^2 |V_{ud}|^2 -|V_{cd}|^2 |V_{us}|^2),\label{Eq:cptnumerator2}
\end{align}
we define the follow asymmetry
\begin{align}
{\mathcal A}_{CPT}^{K_{S,L}^0}&(B_c^{\pm}\rightarrow B^{\pm}K_{S,L}^0)\nonumber\\
&=\frac{\left[\Gamma(B_c^{-}\rightarrow  B^- K_S^0)-\Gamma(B_c^{+}\rightarrow  B^+ K_S^0)\right]-\left[\Gamma(B_c^{-}\rightarrow  B^- K_L^0)-\Gamma(B_c^{+}\rightarrow  B^+ K_L^0)\right]}{\Gamma(B_c^{-}\rightarrow  B^- K_S^0)+\Gamma(B_c^{+}\rightarrow  B^+ K_S^0)+\Gamma(B_c^{-}\rightarrow  B^- K_L^0)+\Gamma(B_c^{+}\rightarrow  B^+ K_L^0)}.\label{Eq:cptasymmetry1}
\end{align}
Substituting Eqs.(\ref{Eq:decaywidth1}), (\ref{Eq:decaywidth2}), (\ref{Eq:decaywidth3}), (\ref{Eq:decaywidth4}), (\ref{Eq:cptnumerator1}) and (\ref{Eq:cptnumerator2}) into Eq.(\ref{Eq:cptasymmetry1}), we can obtain
\begin{align}
{\mathcal A}_{CPT}^{K_{S,L}^0}(B_c^{\pm}\rightarrow B^{\pm}K_{S,L}^0)=&-\frac{Re(z)}{(1+\frac{|z|^2}{4})}\cdot\frac{|V_{cs}|^2 |V_{ud}|^2 -|V_{cd}|^2 |V_{us}|^2}{|V_{cs}|^2 |V_{ud}|^2 +|V_{cd}|^2 |V_{us}|^2}\nonumber\\
& +Re\left(t_{K^0_S-K_L^0}\cdot\left(\frac{p-q }{p+q }+\frac{z}{2}\right)\right)\cdot\frac{|V_{cs}|^2 |V_{ud}|^2 -|V_{cd}|^2 |V_{us}|^2}{|V_{cs}|^2 |V_{ud}|^2 +|V_{cd}|^2 |V_{us}|^2}\nonumber\\
&+\frac{|q|^2  -|p|^2}{|q|^2 +|p|^2}\cdot\frac{|V_{cs}|^2 |V_{ud}|^2 -|V_{cd}|^2 |V_{us}|^2}{|V_{cs}|^2 |V_{ud}|^2 +|V_{cd}|^2 |V_{us}|^2}\nonumber\\
&~~\cdot\frac{{|f_0(m_S^2)|}^2f(m_{B_c^+},m_{B^+},m_S)-{|f_0(m_L^2)|}^2f(m_{B_c^+},m_{B^+} ,m_L)}{{|f_0(m_S^2)|}^2 f(m_{B_c^+},m_{B^+},m_S)+{|f_0(m_L^2)|}^2f(m_{B_c^+},m_{B^+} ,m_L)},\label{Eq:cptasymmetry2}
\end{align}
here, $(|q|^2  -|p|^2)/(|q|^2 +|p|^2)$ is the mixing-induced CP violation, which is of the order of $10^{-3}$~\cite{Lipkin:1999qz,Zyla:2020zbs}. $({|f_0(m_S^2)|}^2f(m_{B_c^+},m_{B^+},m_S)-{|f_0(m_L^2)|}^2f(m_{B_c^+},m_{B^+} ,m_L))/({|f_0(m_S^2)|}^2 f(m_{B_c^+},m_{B^+},m_S)+{|f_0(m_L^2)|}^2f(m_{B_c^+},m_{B^+} ,m_L))$ denotes the result of the $K_S^0-K_L^0$ mass difference and is at the order of $10^{-16}$, which can be calculated by Eq.(\ref{Eq:termofkslmdif}) and its result, so the third term on the right side of Eq.(\ref{Eq:cptasymmetry2}) is immeasurably small (on the order of $10^{-19}$), we can safely neglect this term. In this approach, we obtain
\begin{align}
{\mathcal A}_{CPT}^{K_{S,L}^0}(B_c^{\pm}\rightarrow B^{\pm}K_{S,L}^0)=&-\frac{Re(z)}{(1+\frac{|z|^2}{4})}\cdot\frac{|V_{cs}|^2 |V_{ud}|^2 -|V_{cd}|^2 |V_{us}|^2}{|V_{cs}|^2 |V_{ud}|^2 +|V_{cd}|^2 |V_{us}|^2}\nonumber\\
& +Re\left(t_{K^0_S-K_L^0}\cdot\left(\frac{p-q }{p+q }+\frac{z}{2}\right)\right)\cdot\frac{|V_{cs}|^2 |V_{ud}|^2 -|V_{cd}|^2 |V_{us}|^2}{|V_{cs}|^2 |V_{ud}|^2 +|V_{cd}|^2 |V_{us}|^2}.\label{Eq:cptasymmetryjh}
\end{align}
Obviously, ${\mathcal A}_{CPT}^{K_{S,L}^0}(B_c^{\pm}\rightarrow B^{\pm}K_{S,L}^0)$ contains the term $Re\left(t_{K^0_S-K_L^0}\cdot\frac{p-q }{p+q }\right)$, which is independent from the CPT violation parameter $z$, so the precise calculation of $Re\left(t_{K^0_S-K_L^0}\cdot\frac{p-q }{p+q }\right)$, which is the function of the parameters $m_L-m_S$, $\Gamma_L$, $\Gamma_S$, $p$, $q$, $t_0$ and $t_1$, is crucial to constraint the CPT violation parameter $z$ in the $B_c^{\pm}\rightarrow B^{\pm}K_{S,L}^0$ decays. Here, we note that the values of $t_0$ and $t_1$ must be consistent with the event selection criteria in experiment. In a word, ${\mathcal A}_{CPT}^{K_{S,L}^0}(B_c^{\pm}\rightarrow B^{\pm}K_{S,L}^0)$ can in principle constraint the parameter of CPT violation only if the measurements of the parameters $m_L-m_S$, $\Gamma_L$, $\Gamma_S$, $p$ and $q$ have high precision and the values of $t_0$ and $t_1$ are consistent with the event selection criteria in experiment.

In experiment, the intermediate $B^{\pm}$ states are reconstructed in the final states $\bar{f}_{B^-}$ or $f_{B^-}$, so we defined the following observable
\begin{align}
&{\mathcal A}_{CPT}=\frac{{\mathcal A}_{CPT}^{K_S^0,-}-{\mathcal A}_{CPT}^{K_L^0,-}}{{\mathcal A}_{CPT}^{K_S^0,+}+{\mathcal A}_{CPT}^{K_L^0,+}},\label{Eq:cptasyobserve1}
\end{align}
where ${\mathcal A}_{CPT}^{K_S^0,\pm}$ and ${\mathcal A}_{CPT}^{K_L^0,\pm}$ are defined as
\begin{align}
{\mathcal A}_{CPT}^{K_S^0,\pm}=\Gamma(B_c^{-}\rightarrow  B^- K_S^0)\Gamma(B^{-}\rightarrow f_{B^-})\pm \Gamma(B_c^{+}\rightarrow  B^+ K_S^0)\Gamma(B^{+}\rightarrow \bar{f}_{B^-}),\label{Eq:cptasyobserve2}\\
{\mathcal A}_{CPT}^{K_L^0,\pm}=\Gamma(B_c^{-}\rightarrow  B^- K_L^0)\Gamma(B^{-}\rightarrow f_{B^-})\pm \Gamma(B_c^{+}\rightarrow  B^+ K_L^0)\Gamma(B^{+}\rightarrow \bar{f}_{B^-}).\label{Eq:cptasyobserve3}
\end{align}
By using Eqs.(\ref{Eq:cptasyobserre1}-\ref{Eq:cptasyobsertot}) listed in Appendix~\ref{sec:appendixcptcal}, we can obtain
\begin{align}
{\mathcal A}_{CPT}&=R\left(B_c^{+}\rightarrow  B^+ K_{S,L}^0 \right) \cdot{\mathcal A}_{CP}\left(B^-\rightarrow f_{B^-}\right)+{\mathcal A}_{CPT}^{K_{S,L}^0}(B_c^{\pm}\rightarrow B^{\pm}K_{S,L}^0))\nonumber\\
&=\frac{\Gamma(B_c^{+}\rightarrow  B^+ K_S^0)-\Gamma(B_c^{+}\rightarrow  B^+ K_L^0)}{\Gamma(B_c^{+}\rightarrow  B^+ K_S^0)+\Gamma(B_c^{+}\rightarrow  B^+ K_L^0)}\cdot\frac{\Gamma(B^{-}\rightarrow f_{B^-})-\Gamma(B^{+}\rightarrow \bar{f}_{B^-})}{\Gamma(B^{-}\rightarrow f_{B^-})+\Gamma(B^{+}\rightarrow \bar{f}_{B^-})}\nonumber\\
&+\frac{\left[\Gamma(B_c^{-}\rightarrow  B^- K_S^0)-\Gamma(B_c^{+}\rightarrow  B^+ K_S^0)\right]-\left[\Gamma(B_c^{-}\rightarrow  B^- K_L^0)-\Gamma(B_c^{+}\rightarrow  B^+ K_L^0)\right]}{\Gamma(B_c^{-}\rightarrow  B^- K_S^0)+\Gamma(B_c^{+}\rightarrow  B^+ K_S^0)+\Gamma(B_c^{-}\rightarrow  B^- K_L^0)+\Gamma(B_c^{+}\rightarrow  B^+ K_L^0)}.\label{Eq:cptasyobserre7}
\end{align}
For the semileptonic decays of  $B^{\pm}$ to charm meson, ${\mathcal A}_{CP}\left(B^-\rightarrow f_{B^-}\right)$ is zero in the SM, so ${\mathcal A}_{CPT}$ can in principle constraint the parameter of CPT violation in the semileptonic decays of  $B^{\pm}$ to charm meson, provided the new physics effects are negligible small. By combining Eq.(\ref{Eq:cptasymmetryjh}) and Eq.(\ref{Eq:cptasyobserre7}), we can obtain
\begin{align}
&{\mathcal A}_{CPT}\left(B_c^{+}\rightarrow  B^+ K_{S,L}^0 \rightarrow \bar{D}^0 (\bar{D}^* (2007)^0)  l^+ \nu_l K_{S,L}^0\right)=-\frac{Re(z)}{(1+\frac{|z|^2}{4})}\cdot\frac{|V_{cs}|^2 |V_{ud}|^2 -|V_{cd}|^2 |V_{us}|^2}{|V_{cs}|^2 |V_{ud}|^2 +|V_{cd}|^2 |V_{us}|^2}\nonumber\\
&~~~~~~~~~~~~~~~~~~~~~~~~~~~~~~~~~~~~~+Re\left(t_{K^0_S-K_L^0}\cdot\left(\frac{p-q }{p+q }+\frac{z}{2}\right)\right)\cdot\frac{|V_{cs}|^2 |V_{ud}|^2 -|V_{cd}|^2 |V_{us}|^2}{|V_{cs}|^2 |V_{ud}|^2 +|V_{cd}|^2 |V_{us}|^2}.\label{Eq:cptasyobserre8}
\end{align}
For the nonleptonic decays of  $B^{\pm}$, ${\mathcal A}_{CP}\left(B^-\rightarrow f_{B^-}\right)$ is nonzero, and what's more, the values of $R\left(B_c^{+}\rightarrow  B^+ K_{S,L}^0 \right)$ have not been precisely measured up to now, so it is difficult to constraint the CPT violating parameter $z$ by measuring the value of ${\mathcal A}_{CPT}$ via the nonleptonic decays of the $B^{\pm}$ mesons.
\section{Numerical calculation}
\label{sec:numcalution}
Now, we are going to calculate the values of the observables. The input parameters used in this paper are collected as below~\cite{Zyla:2020zbs,Bailey:2018feb,ExtendedTwistedMass:2020tvp}
\begin{align}
&m_{B_c^+}=6.275\text{GeV}, &m_{B^+}&=5.279 \text{GeV},\nonumber\\
&m_{S}=0.498\text{GeV},  &m_{L}&=0.498 \text{GeV},\nonumber\\
&\Delta m=(3.48\pm0.01)\times 10^{-15}\text{GeV},  &G_F&=1.166\times 10^{-5} \text{GeV}^{-2},\nonumber\\
&\Gamma_S=7.35\times 10^{-15}\text{GeV},  &\Gamma_L&=(1.29\pm0.01)\times 10^{-17} \text{GeV},\label{Eq:parametervalue}\\
&m_{D^\ast}=2.007\text{GeV},  &m_{D_0^\ast}&=2.300\text{GeV},\nonumber\\
&\tau_{B_c}=(0.510\pm0.009)\times 10^{-12}\text{s},  &f_K&=(0.154\pm0.002)\text{GeV},\nonumber\\
&Re(\epsilon)=(1.66\pm0.02)\times 10^{-3},  &Im(\epsilon)&=(1.57\pm0.02)\times 10^{-3}.\nonumber
\end{align}
We use the Wolfenstein parametrization to write the CKM matrix elements~\cite{Zyla:2020zbs,Buras:1998raa}
\begin{align}
&V_{ud}=\sqrt{1-\lambda^2}\sqrt{1-A^2\lambda^6(\rho^2+\eta^2)}, ~~~ &V_{cd}&=-\lambda\sqrt{1-A^2\lambda^4}-A^2\lambda^5\sqrt{1-\lambda^2}(\rho+i\eta),\nonumber\\
&V_{us}=\lambda\sqrt{1-A^2\lambda^6(\rho^2+\eta^2)}, ~~~ &V_{cs}&=\sqrt{1-\lambda^2}\sqrt{1-A^2\lambda^4}-A^2\lambda^6(\rho+i\eta),\label{Eq:ckmwolfenpar}
\end{align}
where $\lambda$, $A$, $\rho$ and $\eta$ are the real parameters. The latest results fitted by the UTfit collaboration~\cite{Ref:utfit} are presented as following
\begin{align}
&\lambda=0.225\pm 0.001, ~~~ &A&=0.826\pm0.012,~~~ &\rho&=0.152\pm0.014,~~~ &\eta&=0.357\pm0.010.\label{Eq:ckmwolfenparval}
\end{align}
With all these parameters and Eqs.(\ref{Eq:decaywidth1}), (\ref{Eq:decaywidth2}), (\ref{Eq:decaywidth3}) and (\ref{Eq:decaywidth4}), we obtain the branching fractions of the $B_c^{\pm}\rightarrow  B^{\pm} K_{S,L}^0$ decays
\begin{align}
&{\mathcal B}(B_c^{\pm}\rightarrow B^{\pm} K_{S}^0)=(2.05^{-0.20}_{+0.21})\times 10^{-3},\label{Eq:bctobksbranch}\\
&{\mathcal B}(B_c^{\pm}\rightarrow B^{\pm} K_{L}^0)=(2.54^{-0.25}_{+0.26})\times 10^{-3},\label{Eq:bctobklbranch}
\end{align}
here, the above results are the averaged branching ratios of the decay and its charge conjugate. By using the branching ratios of the consequent decays of the $B^{\pm}$ and $K_{S,L}^0$ mesons given by the Particle Data Group, we can calculate the results of ${\mathcal B}(B_c^{\pm}\rightarrow B^{\pm} K_{S,L}^0\rightarrow f_{B^\pm} f_{K_{S,L}^0})$, which are listed in Table~\ref{totbranchratio}.
\begin{table}[t]
\begin{center}
\caption{\label{totbranchratio} \small The branching fractions of the $B_c^{\pm}\rightarrow B^{\pm} K_{S,L}^0\rightarrow f_{B^\pm} f_{K_{S,L}^0}$ decays.}
\vspace{0.1cm}
\doublerulesep 0.8pt \tabcolsep 0.18in
\scriptsize
\begin{tabular}{c|c}
\hline
the decay channel & the branching ratio \\
\hline
${\mathcal B}(B_c^+ \rightarrow B^{+} K_{S}^0\rightarrow\bar{D}^0 l^{+}\nu_l K_{S}^0\rightarrow K^+ e^-\bar{\nu}_e l^{+}\nu_l \pi^+\pi^- )$ & $\left(1.18^{-0.12}_{+0.13}\right)\times 10^{-6}$   \\
\hline
${\mathcal B}(B_c^+ \rightarrow B^{+} K_{S}^0\rightarrow\bar{D}^0 l^{+}\nu_l K_{S}^0\rightarrow K^+ \pi^- l^{+}\nu_l \pi^+\pi^- )$ & $\left(1.32\pm 0.14\right)\times 10^{-6}$   \\
\hline
${\mathcal B}(B_c^+ \rightarrow B^{+} K_{S}^0\rightarrow\bar{D}^0 l^{+}\nu_l K_{S}^0\rightarrow K^+ \pi^- \pi^0 l^{+}\nu_l \pi^+\pi^- )$ & $\left(4.80_{+ 0.55}^{-0.53}\right)\times 10^{-6}$   \\
\hline
${\mathcal B}(B_c^+ \rightarrow B^{+} K_{S}^0\rightarrow\bar{D}^* (2007)^0 l^{+}\nu_l K_{S}^0\rightarrow\bar{D}^0 \pi^0 l^{+}\nu_l K_{S}^0 \rightarrow K^+  e^-\bar{\nu}_e \pi^0 l^{+}\nu_l \pi^+\pi^-  )$ & $\left(1.84\pm 0.20\right)\times 10^{-6}$   \\
\hline
${\mathcal B}(B_c^+ \rightarrow B^{+} K_{S}^0\rightarrow\bar{D}^* (2007)^0 l^{+}\nu_l K_{S}^0\rightarrow\bar{D}^0 \pi^0 l^{+}\nu_l K_{S}^0 \rightarrow K^+  \pi^- \pi^0 l^{+}\nu_l \pi^+\pi^-  )$ & $\left(2.05^{-0.22}_{+0.23}\right)\times 10^{-6}$   \\
\hline
${\mathcal B}(B_c^+ \rightarrow B^{+} K_{S}^0\rightarrow\bar{D}^* (2007)^0 l^{+}\nu_l K_{S}^0\rightarrow\bar{D}^0 \pi^0 l^{+}\nu_l K_{S}^0 \rightarrow K^+  \pi^- \pi^0\pi^0 l^{+}\nu_l \pi^+\pi^-  )$ & $\left(7.48^{-0.83}_{+0.87}\right)\times 10^{-6}$   \\
\hline
${\mathcal B}(B_c^+ \rightarrow B^{+} K_{S}^0\rightarrow\bar{D}^0 \pi^{+} K_{S}^0\rightarrow K^+ \pi^- \pi^0 \pi^{+}\pi^+\pi^- )$ & $\left(9.56^{-1.03}_{+1.07}\right)\times 10^{-7}$   \\
\hline
${\mathcal B}(B_c^+ \rightarrow B^{+} K_{S}^0\rightarrow\bar{D}^0 \rho^{+} K_{S}^0\rightarrow  K^+ e^-\bar{\nu}_e  \pi^{+} \pi^0 \pi^+\pi^- )$ & $\left(6.73^{-1.12}_{+1.14}\right)\times 10^{-7}$   \\
\hline
${\mathcal B}(B_c^+ \rightarrow B^{+} K_{S}^0\rightarrow\bar{D}^0 \rho^{+} K_{S}^0\rightarrow  K^+ \pi^-  \pi^{+} \pi^0 \pi^+\pi^- )$ & $\left(7.51^{-1.25}_{+1.27}\right)\times 10^{-7}$   \\
\hline
${\mathcal B}(B_c^+ \rightarrow B^{+} K_{S}^0\rightarrow\bar{D}^0 \rho^{+} K_{S}^0\rightarrow  K^+ \pi^- \pi^0  \pi^{+} \pi^0 \pi^+\pi^- )$ & $\left(2.74^{-0.46}_{+0.47}\right)\times 10^{-6}$   \\
\hline
${\mathcal B}(B_c^+ \rightarrow B^{+} K_{S}^0\rightarrow\bar{D}^* (2007)^0 \pi^{+} K_{S}^0\rightarrow\bar{D}^0 \pi^0 \pi^{+} K_{S}^0 \rightarrow K^+  \pi^- \pi^0\pi^0 \pi^{+} \pi^+\pi^-  )$ & $\left(6.48^{-0.71}_{+0.74}\right)\times 10^{-7}$   \\
\hline
${\mathcal B}(B_c^+ \rightarrow B^{+} K_{S}^0\rightarrow\bar{D}^* (2007)^0 \pi^{+}\pi^+\pi^- K_{S}^0\rightarrow\bar{D}^0 \pi^0 \pi^{+}\pi^+\pi^- K_{S}^0 \rightarrow K^+  \pi^- \pi^0 \pi^0 \pi^{+} \pi^+\pi^-  \pi^+\pi^-  )$ & $\left(1.36^{-0.21}_{+0.22}\right)\times 10^{-6}$   \\
\hline
${\mathcal B}(B_c^+ \rightarrow B^{+} K_{S}^0\rightarrow J/\psi K^{+}K_{S}^0\rightarrow e^{+} e^- K^+   \pi^+\pi^-  )$ & $\left(8.52^{-0.86}_{+0.90}\right)\times 10^{-8}$   \\
\hline
${\mathcal B}(B_c^+ \rightarrow B^{+} K_{L}^0\rightarrow\bar{D}^0 l^{+}\nu_l K_{L}^0\rightarrow K^+ e^-\bar{\nu}_e l^{+}\nu_l \pi^\pm e^\mp \nu_e )$ & $\left(8.57^{-0.91}_{+0.94}\right)\times 10^{-7}$   \\
\hline
${\mathcal B}(B_c^+ \rightarrow B^{+} K_{L}^0\rightarrow\bar{D}^0 l^{+}\nu_l K_{L}^0\rightarrow K^+ \pi^- l^{+}\nu_l \pi^\pm e^\mp \nu_e )$ & $\left(9.56^{-1.01}_{+1.05}\right)\times 10^{-7}$   \\
\hline
${\mathcal B}(B_c^+ \rightarrow B^{+} K_{L}^0\rightarrow\bar{D}^0 l^{+}\nu_l K_{L}^0\rightarrow K^+ \pi^- \pi^0 l^{+}\nu_l \pi^\pm e^\mp \nu_e )$ & $\left(3.49^{-0.39}_{+0.40}\right)\times 10^{-6}$   \\
\hline
${\mathcal B}(B_c^+ \rightarrow B^{+} K_{L}^0\rightarrow\bar{D}^0 l^{+}\nu_l K_{L}^0\rightarrow K^+ \pi^- \pi^0 l^{+}\nu_l \pi^+ \pi^- \pi^0 )$ & $\left(1.08\pm 0.12\right)\times 10^{-6}$   \\
\hline
${\mathcal B}(B_c^+ \rightarrow B^{+} K_{L}^0\rightarrow\bar{D}^* (2007)^0 l^{+}\nu_l K_{L}^0\rightarrow\bar{D}^0 \pi^0 l^{+}\nu_l K_{L}^0 \rightarrow K^+  e^-\bar{\nu}_e \pi^0 l^{+}\nu_l \pi^\pm e^\mp \nu_e  )$ & $\left(1.34^{-0.14}_{+0.15} \right)\times 10^{-6}$   \\
\hline
${\mathcal B}(B_c^+ \rightarrow B^{+} K_{L}^0\rightarrow\bar{D}^* (2007)^0 l^{+}\nu_l K_{L}^0\rightarrow\bar{D}^0 \pi^0 l^{+}\nu_l K_{L}^0 \rightarrow K^+  \pi^- \pi^0 l^{+}\nu_l \pi^\pm e^\mp \nu_e  )$ & $\left(1.49\pm 0.16 \right)\times 10^{-6}$   \\
\hline
${\mathcal B}(B_c^+ \rightarrow B^{+} K_{L}^0\rightarrow\bar{D}^* (2007)^0 l^{+}\nu_l K_{L}^0\rightarrow\bar{D}^0 \pi^0 l^{+}\nu_l K_{L}^0 \rightarrow K^+  \pi^- \pi^0 \pi^0 l^{+}\nu_l \pi^\pm e^\mp \nu_e  )$ & $\left(5.43^{-0.61}_{+0.63}\right)\times 10^{-6}$   \\
\hline
${\mathcal B}(B_c^+ \rightarrow B^{+} K_{L}^0\rightarrow\bar{D}^* (2007)^0 l^{+}\nu_l K_{L}^0\rightarrow\bar{D}^0 \pi^0 l^{+}\nu_l K_{L}^0 \rightarrow K^+  \pi^- \pi^0 \pi^0 l^{+}\nu_l \pi^+ \pi^- \pi^0  )$ & $\left(1.68\pm 0.19\right)\times 10^{-6}$   \\
\hline
${\mathcal B}(B_c^+ \rightarrow B^{+} K_{L}^0\rightarrow\bar{D}^0 \pi^{+} K_{L}^0\rightarrow K^+ \pi^- \pi^0 \pi^{+}\pi^\pm e^\mp \nu_e)$ & $\left(6.94^{-0.75}_{+0.77}\right)\times 10^{-7}$   \\
\hline
${\mathcal B}(B_c^+ \rightarrow B^{+} K_{L}^0\rightarrow\bar{D}^0 \rho^{+} K_{L}^0\rightarrow  K^+ \pi^-  \pi^{+} \pi^0 \pi^\pm e^\mp \nu_e )$ & $\left(5.45^{-0.91}_{+0.92}\right)\times 10^{-7}$   \\
\hline
${\mathcal B}(B_c^+ \rightarrow B^{+} K_{L}^0\rightarrow\bar{D}^0 \rho^{+} K_{L}^0\rightarrow  K^+ \pi^- \pi^0 \pi^{+} \pi^0 \pi^\pm e^\mp \nu_e )$ & $\left(1.99\pm0.34\right)\times 10^{-6}$   \\
\hline
${\mathcal B}(B_c^+ \rightarrow B^{+} K_{L}^0\rightarrow\bar{D}^0 \rho^{+} K_{L}^0\rightarrow  K^+ \pi^- \pi^0 \pi^{+} \pi^0 \pi^+ \pi^- \pi^0 )$ & $\left(6.15^{-1.05}_{+1.06}\right)\times 10^{-7}$   \\
\hline
${\mathcal B}(B_c^+ \rightarrow B^{+} K_{L}^0\rightarrow\bar{D}^* (2007)^0 \pi^{+}\pi^+\pi^- K_{L}^0\rightarrow\bar{D}^0 \pi^0 \pi^{+}\pi^+\pi^- K_{L}^0 \rightarrow K^+  \pi^- \pi^0 \pi^0 \pi^{+} \pi^+\pi^-  \pi^\pm e^\mp \nu_e  )$ & $\left(9.88^{-1.55}_{+1.57}\right)\times 10^{-7}$   \\
\hline
${\mathcal B}(B_c^+ \rightarrow B^{+} K_{L}^0\rightarrow J/\psi K^{+}K_{L}^0\rightarrow e^{+} e^- K^+   \pi^\pm e^\mp \nu_e  )$ & $\left(6.19^{-0.63}_{+0.66}\right)\times 10^{-8}$   \\
\hline
\end{tabular}
\end{center}
\end{table}
Here, we only consider the decay channels with the branching ratios are larger than $5.0\times 10^{-7}$, which are hopefully to be marginally observed by the current experiments. We also list the $B_c^+ \rightarrow B^{+} K_{S,L}^0\rightarrow J/\psi K^{+}K_{S,L}^0\rightarrow e^{+} e^- K^+  \pi^+\pi^- ( \pi^\pm e^\mp \nu_e) $ decays in Table~\ref{totbranchratio}, because these decays have a small number of final states and can be easily detected.

By substituting the values of the parameters in Eqs.(\ref{Eq:parametervalue}), (\ref{Eq:ckmwolfenpar}) and (\ref{Eq:ckmwolfenparval}) into Eqs.(\ref{Eq:cpbcpmtobpmks}) and (\ref{Eq:cpbcpmtobpmkl}), we can proceed to calculate the numerical results of the CP asymmetries in $B_c^{\pm}\rightarrow B^{\pm} K_{S,L}^0$ decays
\begin{align}
&{\mathcal A}_{CP}\left(B_c^\pm\rightarrow  B^{\pm}K_{S}^0\right)=\frac{\Gamma(B_c^-\rightarrow B^{-}K_{S}^0)-\Gamma(B_c^+\rightarrow B^{+}K_{S}^0)}{\Gamma(B_c^-\rightarrow B^{-}K_{S}^0)+\Gamma(B_c^+\rightarrow B^{+}K_{S}^0)}=(4.05\pm 0.05)\times 10^{-3},\label{Eq:numcpasyks}\\
&{\mathcal A}_{CP}\left(B_c^\pm\rightarrow  B^{\pm}K_{L}^0\right)=\frac{\Gamma(B_c^-\rightarrow B^{-}K_{L}^0)-\Gamma(B_c^+\rightarrow B^{+}K_{L}^0)}{\Gamma(B_c^-\rightarrow B^{-}K_{L}^0)+\Gamma(B_c^+\rightarrow B^{+}K_{L}^0)}=(-2.98\pm 0.04)\times 10^{-3}.\label{Eq:numcpasykl}
\end{align}
Combining Eqs.(\ref{Eq:cpasymmsimksl}), (\ref{Eq:cpbplusdzlvl}), (\ref{Eq:numcpasyks}) and (\ref{Eq:numcpasykl}), we can obtain the time-independent CP asymmetries ${\mathcal A}_{CP}^{K_{S}^0}$ and ${\mathcal A}_{CP}^{K_{L}^0}$ in the $B_c^{\pm}\rightarrow B^{\pm} K_{S,L}^0\rightarrow f_{B^{-}}(\bar{f}_{B^{-}}) f_{K_{S,L}^0}$ decays. As for the $B^{\pm}\rightarrow \bar{D}^0 (D^0) l^{\pm} \nu_{l}$ and $B^{\pm}\rightarrow \bar{D}^* (2007)^0 (D^*(2007)^0) l^{\pm} \nu_{l}$ decays, the CP asymmetries in these decays are zero in the SM, so we can obtain
\begin{align}
&{\mathcal A}_{CP}^{K_{S}^0}\left(B_c^+\rightarrow  B^{+}K_{S}^0\rightarrow \bar{D}^0 l^+\nu_l K_{S}^0 \right)=(4.05\pm 0.05)\times 10^{-3}+\frac{\Gamma(D^0\rightarrow f_{D})-\Gamma(\bar{D}^0\rightarrow \bar{f}_{D})}{\Gamma(D^0\rightarrow f_{D})+\Gamma(\bar{D}^0\rightarrow \bar{f}_{D})},\label{Eq:numcpasykstotdz}\\
&{\mathcal A}_{CP}^{K_{L}^0}\left(B_c^+\rightarrow  B^{+}K_{L}^0\rightarrow \bar{D}^0 l^+\nu_l K_{L}^0 \right)=(-2.98\pm 0.04)\times 10^{-3}+\frac{\Gamma(D^0\rightarrow f_{D})-\Gamma(\bar{D}^0\rightarrow \bar{f}_{D})}{\Gamma(D^0\rightarrow f_{D})+\Gamma(\bar{D}^0\rightarrow \bar{f}_{D})},\label{Eq:numcpasykltotdz}\\
&{\mathcal A}_{CP}^{K_{S}^0}\left(B_c^+\rightarrow  B^{+}K_{S}^0\rightarrow \bar{D}^*(2007)^0 l^+\nu_l K_{S}^0 \rightarrow \bar{D}^0 \pi^0 l^+\nu_l K_{S}^0\right)\nonumber \\
&~~~~~~~~~~~~~~~~~~~~~~~~~~~~~~~~~~~~~~~~~=(4.05\pm 0.05)\times 10^{-3}+\frac{\Gamma(D^0\rightarrow f_{D})-\Gamma(\bar{D}^0\rightarrow \bar{f}_{D})}{\Gamma(D^0\rightarrow f_{D})+\Gamma(\bar{D}^0\rightarrow \bar{f}_{D})},\label{Eq:numcpasykstotdstar}\\
&{\mathcal A}_{CP}^{K_{L}^0}\left(B_c^+\rightarrow  B^{+}K_{L}^0\rightarrow \bar{D}^*(2007)^0 l^+\nu_l K_{L}^0 \rightarrow \bar{D}^0 \pi^0 l^+\nu_l K_{L}^0\right)\nonumber \\
&~~~~~~~~~~~~~~~~~~~~~~~~~~~~~~~~~~~~~~~~~=(-2.98\pm 0.04)\times 10^{-3}+\frac{\Gamma(D^0\rightarrow f_{D})-\Gamma(\bar{D}^0\rightarrow \bar{f}_{D})}{\Gamma(D^0\rightarrow f_{D})+\Gamma(\bar{D}^0\rightarrow \bar{f}_{D})},\label{Eq:numcpasykltotdz}
\end{align}
of course, $(\Gamma(D^0\rightarrow f_{D})-\Gamma(\bar{D}^0\rightarrow \bar{f}_{D}))/(\Gamma(D^0\rightarrow f_{D})+\Gamma(\bar{D}^0\rightarrow \bar{f}_{D}))$ is zero in the semileptonic decay of the $D^0$ meson in the SM. Meanwhile, the CP violations in the nonleptonic decays of charm mesons, such as $D^0\rightarrow K^-\pi^+$ and $D^0\rightarrow K^-\pi^+ \pi^0$, are of order of $10^{-4}$, or even smaller~\cite{Zyla:2020zbs,Bigi:1994aw,Bianco:2003vb,Petrov:2011un,LHCb:2013zpr,LHCb:2016qsa}, so the CP violations in the nonleptonic decays of charm mesons are smaller than that in $B_c^\pm \rightarrow B^{\pm} K_{S,L}^0$ decays and can be safely neglected. From Table~\ref{totbranchratio}, we can see that the branching ratios of the following decay channels are of the order of $10^{-6}$: $B_c^+ \rightarrow B^{+} K_{S}^0\rightarrow\bar{D}^0 l^{+}\nu_l K_{S}^0\rightarrow K^+ e^-\bar{\nu}_e l^{+}\nu_l \pi^+\pi^-$, $B_c^+ \rightarrow B^{+} K_{S}^0\rightarrow\bar{D}^0 l^{+}\nu_l K_{S}^0\rightarrow K^+ \pi^- l^{+}\nu_l \pi^+\pi^-$, $B_c^+ \rightarrow B^{+} K_{S}^0\rightarrow\bar{D}^0 l^{+}\nu_l K_{S}^0\rightarrow K^+ \pi^- \pi^0 l^{+}\nu_l \pi^+\pi^-$, $B_c^+ \rightarrow B^{+} K_{S}^0\rightarrow\bar{D}^* (2007)^0 l^{+}\nu_l K_{S}^0\rightarrow\bar{D}^0 \pi^0 l^{+}\nu_l K_{S}^0 \rightarrow K^+  e^-\bar{\nu}_e \pi^0 l^{+}\nu_l \pi^+\pi^-$, $B_c^+ \rightarrow B^{+} K_{S}^0\rightarrow\bar{D}^* (2007)^0 l^{+}\nu_l K_{S}^0\rightarrow\bar{D}^0 \pi^0 l^{+}\nu_l K_{S}^0 \rightarrow K^+  \pi^- \pi^0 l^{+}\nu_l \pi^+\pi^-$, $B_c^+ \rightarrow B^{+} K_{S}^0\rightarrow\bar{D}^* (2007)^0 l^{+}\nu_l K_{S}^0\rightarrow\bar{D}^0 \pi^0 l^{+}\nu_l K_{S}^0 \rightarrow K^+  \pi^- \pi^0\pi^0 l^{+}\nu_l \pi^+\pi^-$, $B_c^+ \rightarrow B^{+} K_{L}^0\rightarrow\bar{D}^0 l^{+}\nu_l K_{L}^0\rightarrow K^+ \pi^- \pi^0 l^{+}\nu_l \pi^\pm e^\mp \nu_e$, $B_c^+ \rightarrow B^{+} K_{L}^0\rightarrow\bar{D}^0 l^{+}\nu_l K_{L}^0\rightarrow K^+ \pi^- \pi^0 l^{+}\nu_l$ $\pi^+ \pi^- \pi^0$, $B_c^+ \rightarrow B^{+} K_{L}^0\rightarrow\bar{D}^* (2007)^0 l^{+}\nu_l K_{L}^0\rightarrow\bar{D}^0 \pi^0 l^{+}\nu_l K_{L}^0 \rightarrow K^+  e^-\bar{\nu}_e \pi^0 l^{+}\nu_l \pi^\pm e^\mp \nu_e$, $B_c^+ \rightarrow B^{+} K_{L}^0\rightarrow\bar{D}^* (2007)^0 l^{+}\nu_l K_{L}^0\rightarrow\bar{D}^0 \pi^0 l^{+}\nu_l K_{L}^0 \rightarrow K^+  \pi^- \pi^0 l^{+}\nu_l \pi^\pm e^\mp \nu_e$, $B_c^+ \rightarrow B^{+} K_{L}^0\rightarrow\bar{D}^* (2007)^0 $ $l^{+}\nu_l K_{L}^0\rightarrow\bar{D}^0 \pi^0 l^{+}\nu_l K_{L}^0 \rightarrow K^+  \pi^- \pi^0 \pi^0 l^{+}\nu_l \pi^\pm $ $e^\mp \nu_e$ and $B_c^+ \rightarrow B^{+} K_{L}^0\rightarrow\bar{D}^* (2007)^0 l^{+}\nu_l K_{L}^0\rightarrow\bar{D}^0 \pi^0 l^{+}\nu_l K_{L}^0 \rightarrow K^+  \pi^- \pi^0 \pi^0 l^{+}\nu_l \pi^+ \pi^- \pi^0$, so the magnitude of the branching ratio times ${\mathcal A}_{CP}^{K_{S,L}^0}$ can reach $10^{-9}$ or larger for these channels, which are possible to be observed. Here, one can also consider the possibility to sum over the available final states of $D^0$ decays to obtain a statistically significant signal of CP violation.

For the case of the nonleptonic decays of $B^{\pm}$ mesons, ${\mathcal A}_{CP}^{K_{S,L}^0}$ can receive contributions from both the CP asymmetries in $B^{\pm}\rightarrow \bar{f}_{B^-}(f_{B^-})$ decays and the CP asymmetries in $B_c^{\pm}\rightarrow B^{\pm} K_{S,L}^0$ decays, so it is possible to extract the CP asymmetries in $B_c^{\pm}\rightarrow B^{\pm} K_{S,L}^0$ decays in the channels which have the accurate measurements of the CP asymmetries in $B^{\pm}\rightarrow \bar{f}_{B^-}(f_{B^-})$ decays and the large branching ratios of the $B_c^{\pm}\rightarrow B^{\pm} K_{S,L}^0\rightarrow \bar{f}_{B^-}(f_{B^-}) f_{K_{S,L}^0} $ decay chains. However, one can see from Particle Data Group~\cite{Zyla:2020zbs} and Table~\ref{totbranchratio} that some decay channels, such as $B_c^{+}\rightarrow B^{+} K_{S}^0\rightarrow \bar{D}^0 \pi^+\pi^+\pi^- \rightarrow K^+ \pi^- \pi^0 \pi^+ \pi^+\pi^-$ and $B_c^{+}\rightarrow B^{+} K_{S}^0\rightarrow \bar{D}^0 \rho^+\pi^+\pi^- \rightarrow K^+ \pi^- \pi^0 \pi^+ \pi^0 \pi^+\pi^-$, have the large branching ratio (at the level of $10^{-6}$) but an inaccurate value of ${\mathcal A}_{CP}\left(B^-\rightarrow f_{B^-}\right)$. Other decays, such as $B_c^{+}\rightarrow B^{+} K_{S}^0\rightarrow K^+ K^- \pi^+ \pi^+\pi^-$ and $B_c^{+}\rightarrow B^{+} K_{S}^0\rightarrow \pi^+\pi^-\pi^+\pi^+\pi^- $, have the accurate measurement of ${\mathcal A}_{CP}\left(B^-\rightarrow f_{B^-}\right)$ but a small branching ratio, further more,  ${\mathcal A}_{CP}\left(B^-\rightarrow f_{B^-}\right)$ in these channels is at the level of $10^{-2}$ or larger, which dominant the ${\mathcal A}_{CP}^{K_{S,L}^{0}}$, so it is difficult to extract the CP asymmetries in $B_c^\pm \rightarrow B^{\pm} K_{S,L}^0$ decays by measuring the CP violations in the $B_c^\pm \rightarrow B^{\pm} K_{S,L}^0\rightarrow f_{B^\pm} f_{K_{S,L}^0}$ processes with the nonleptonic decays of $B^\pm$.

The explicit expressions for the $K_S^0-K_L^0$ asymmetries $R(B_c^+\rightarrow B^+ K_{S,L}^0)$ and $R(B_c^-\rightarrow B^- K_{S,L}^0)$ have been given in Eqs.(\ref{Eq:ksklasyresult1}) and (\ref{Eq:ksklasyresult2}). Using the values of the parameters in Eqs.(\ref{Eq:parametervalue}), (\ref{Eq:ckmwolfenpar}) and (\ref{Eq:ckmwolfenparval}), the numerical results of $R(B_c^+\rightarrow B^+ K_{S,L}^0)$ and $R(B_c^-\rightarrow B^- K_{S,L}^0)$ can be obtained
\begin{align}
&R(B_c^+\rightarrow B^+ K_{S,L}^0)=-0.109\pm 0.001,\label{Eq:numksklasy1}\\
&R(B_c^-\rightarrow B^- K_{S,L}^0)=-0.103\pm 0.001.\label{Eq:numksklasy2}
\end{align}
Obviously, the large values of the $K_S^0-K_L^0$ asymmetries indicate that there exist a large difference between the branching ratios of $B_c^\pm\rightarrow B^\pm K_{S}^0$ and the branching ratios of $B_c^\pm\rightarrow B^\pm K_{L}^0$. Combining Eqs.(\ref{Eq:numksklasy1}) and (\ref{Eq:numksklasy2}) with the branching ratios of the $B_c^\pm \rightarrow B^{\pm} K_{S,L}^0\rightarrow f_{B^\pm} K_{S,L}^0$ decays from Particle Data Group~\cite{Zyla:2020zbs} and Table~\ref{totbranchratio}, we can see that the maximum order of the branching ratios for the decays $B_c^\pm \rightarrow B^{\pm} K_{S,L}^0\rightarrow f_{B^\pm} K_{S,L}^0$ times $R(B_c^\pm\rightarrow B^\pm K_{S,L}^0)$ can reach $10^{-7}$, so $R(B_c^\pm\rightarrow B^\pm K_{S,L}^0)$ are hopefully to be detected in the current experiments.

Now, we move on to the determination or constraint of the parameter of CPT violation in $B_c^\pm\rightarrow B^\pm K_{S,L}^0\rightarrow f_{B^\pm} f_{K_{S,L}^0}$ decays. As discussed in Section~\ref{sec:cpvcptvkskl}, the most promising process to constraint the parameter of CPT violation are the $B_c^{+}\rightarrow B^{+} K_{S,L}^0\rightarrow \bar{D}^0 l^+ \nu_l K_{S,L}^0 \rightarrow K^+ l^- \bar{\nu}_l l^+ \nu_l K_{S,L}^0 $ and $B_c^{+}\rightarrow B^{+} K_{S,L}^0\rightarrow \bar{D}^* (2007)^0 l^+ \nu_l K_{S,L}^0 \rightarrow \bar{D}^0 \pi^0 l^+ \nu_l K_{S,L}^0 \rightarrow K^+ l^- \bar{\nu}_l  \pi^0  l^+ \nu_l K_{S,L}^0 $ decays. By combining Eq.(\ref{Eq:cptasyobserre8}), Eq.(\ref{Eq:ckmwolfenpar}) and Eq.(\ref{Eq:ckmwolfenparval}), we can obtain
\begin{align}
\frac{|V_{cs}|^2 |V_{ud}|^2 -|V_{cd}|^2 |V_{us}|^2}{|V_{cs}|^2 |V_{ud}|^2 +|V_{cd}|^2 |V_{us}|^2}=0.9943\pm 0.0001.\label{Eq:cptckmfactor}
\end{align}
From Table~\ref{totbranchratio}, we can see that the branching ratios of the $B_c^{+}\rightarrow B^{+} K_{S}^0\rightarrow \bar{D}^0 l^+ \nu_l K_{S}^0 \rightarrow K^+ l^- \bar{\nu}_l l^+ \nu_l \pi^+ \pi^- $, $B_c^{+}\rightarrow B^{+} K_{L}^0\rightarrow \bar{D}^0 l^+ \nu_l K_{L}^0 \rightarrow K^+ l^- \bar{\nu}_l l^+ \nu_l \pi^\pm e^\mp \nu_{e} $, $B_c^{+}\rightarrow B^{+} K_{S}^0\rightarrow \bar{D}^* (2007)^0 $  $l^+ \nu_l K_{S}^0 \rightarrow \bar{D}^0 \pi^0 l^+ \nu_l K_{S}^0 \rightarrow K^+ l^- \bar{\nu}_l \pi^0 l^+ \nu_l \pi^+ \pi^- $ and $B_c^{+}\rightarrow B^{+} K_{L}^0\rightarrow \bar{D}^* (2007)^0 l^+ \nu_l K_{L}^0 \rightarrow \bar{D}^0 \pi^0 l^+ \nu_l K_{L}^0 \rightarrow K^+ l^- \bar{\nu}_l \pi^0 l^+ \nu_l \pi^\pm e^\mp \nu_{e}$ decays can arrive at $1.18\times 10^{-6}$, $8.57\times 10^{-7}$, $1.84\times 10^{-6}$ and $1.34\times 10^{-6}$, respectively, so the sensitivity for the measurement of ${\mathcal A}_{CPT}$ is expected to be at the level of  $10^{-3}$ if we assume the selection efficiency is $10^{-3}$ and the total number of $B_{c}^{\pm}$ events is $10^{12}$ after the LHC run 5~\cite{LHCb:2018roe}. With the values of the parameters in Eqs.(\ref{Eq:parametervalue}), (\ref{Eq:ckmwolfenpar}) and (\ref{Eq:ckmwolfenparval}), we can obtain
\begin{align}
&Re\left(t_{K^0_S-K_L^0}\cdot\frac{p-q }{p+q }\right)\cdot\frac{|V_{cs}|^2 |V_{ud}|^2 -|V_{cd}|^2 |V_{us}|^2}{|V_{cs}|^2 |V_{ud}|^2 +|V_{cd}|^2 |V_{us}|^2}=(3.46\pm 0.03)\times 10^{-3},\label{Eq:cpttiemevvalue}
\end{align}
which accuracy can reach $10^{-4}\sim 10^{-5}$. According to Eq.(\ref{Eq:cptasyobserre8}), we can obtain that the sensitivity for the measurement of the CPT violation parameter $z$ is expected to be at the level of  $10^{-3}$.
\section{Prospects for the measurement at LHC}
\label{sec:prosmeslhc}
At the LHC experiment, around $5\times10^{10}$ $B_c$ mesons per years could be produced~\cite{Xiao:2011zz,Zhou:2020bnm}, therefore, one could expect $5\times10^{11}$ $B_c$ events with 10 years' run and 50$fb^{-1}$ data at LHCb~\cite{LHCb:2012myk}. Moreover, LHCb will accumulate a data sample corresponding to a minimum of 300$fb^{-1}$ in the HL-LHC era~\cite{LHCb:2018roe}, so the total number of $B_c$ events may exceed $3\times10^{12}$ by the end of the HL-LHC era. Basing on these samples, the $B_c$ decay chains with a branching ratio at the level of $10^{-8}-10^{-6}$ can be measured with a good precision.

Following Refs.~\cite{Zhou:2020bnm,Dai:1998hb,Fu:2011tn}, we give an estimation of how many $B_c$ events-times-efficiency are needed to establish the branching ratio, the CP asymmetry and the $K_S^0-K_L^0$ asymmetry to three standard deviations (3$\sigma$). When the decays are observed at three standard deviations (3$\sigma$) level, the numbers of $B_c$ events-times-efficiency needed read
\begin{align}
(\epsilon_f N)_{\mathcal B}=\frac{9}{{\mathcal B}(B_c^\pm \rightarrow B^{\pm} K_{S,L}^0\rightarrow f_{B^\pm} f_{K_{S,L}^0 })},\label{Eq:numneedbr}
\end{align}
where $\epsilon_f $ is the detecting efficiency of the final states. With the numerical results of the branching ratios in Table~\ref{totbranchratio}, we can obtain the numerical results of $(\epsilon_f N)_{\mathcal B}$ for the $B_c^\pm \rightarrow B^{\pm} K_{S,L}^0\rightarrow f_{B^\pm} f_{K_{S,L}^0 }$ channels, as shown in Table~\ref{tabnumbrobe}.
\begin{table}[t]
\begin{center}
\caption{\label{tabnumbrobe} \small The numerical results of $(\epsilon_f N)_{\mathcal B}$ for the $B_c^{\pm}\rightarrow B^{\pm} K_{S,L}^0\rightarrow f_{B^\pm} f_{K_{S,L}^0}$ decays.}
\vspace{0.1cm}
\doublerulesep 0.8pt \tabcolsep 0.18in
\scriptsize
\begin{tabular}{c|c}
\hline
the decay channel & $(\epsilon_f N)_{\mathcal B}$ \\
\hline
$B_c^+ \rightarrow B^{+} K_{S}^0\rightarrow\bar{D}^0 l^{+}\nu_l K_{S}^0\rightarrow K^+ e^-\bar{\nu}_e l^{+}\nu_l \pi^+\pi^- $ & $\left(6.87\sim8.52\right)\times 10^{6}$   \\
\hline
$B_c^+ \rightarrow B^{+} K_{S}^0\rightarrow\bar{D}^0 l^{+}\nu_l K_{S}^0\rightarrow K^+ \pi^- l^{+}\nu_l \pi^+\pi^- $ & $\left(6.16\sim7.64\right)\times10^{6}$   \\
\hline
$B_c^+ \rightarrow B^{+} K_{S}^0\rightarrow\bar{D}^0 l^{+}\nu_l K_{S}^0\rightarrow K^+ \pi^- \pi^0 l^{+}\nu_l \pi^+\pi^- $ & $\left(1.68\sim 2.11\right)\times 10^{6}$   \\
\hline
$B_c^+ \rightarrow B^{+} K_{S}^0\rightarrow\bar{D}^* (2007)^0 l^{+}\nu_l K_{S}^0\rightarrow\bar{D}^0 \pi^0 l^{+}\nu_l K_{S}^0 \rightarrow K^+  e^-\bar{\nu}_e \pi^0 l^{+}\nu_l \pi^+\pi^- $ & $\left(4.40\sim 5.47\right)\times 10^{6}$   \\
\hline
$B_c^+ \rightarrow B^{+} K_{S}^0\rightarrow\bar{D}^* (2007)^0 l^{+}\nu_l K_{S}^0\rightarrow\bar{D}^0 \pi^0 l^{+}\nu_l K_{S}^0 \rightarrow K^+  \pi^- \pi^0 l^{+}\nu_l \pi^+\pi^- $ & $\left(3.95\sim4.91\right)\times 10^{6}$   \\
\hline
$B_c^+ \rightarrow B^{+} K_{S}^0\rightarrow\bar{D}^* (2007)^0 l^{+}\nu_l K_{S}^0\rightarrow\bar{D}^0 \pi^0 l^{+}\nu_l K_{S}^0 \rightarrow K^+  \pi^- \pi^0\pi^0 l^{+}\nu_l \pi^+\pi^-  $ & $\left(1.08\sim1.35\right)\times 10^{6}$   \\
\hline
$B_c^+ \rightarrow B^{+} K_{S}^0\rightarrow\bar{D}^0 \pi^{+} K_{S}^0\rightarrow K^+ \pi^- \pi^0 \pi^{+}\pi^+\pi^- $ & $\left(0.85\sim1.05\right)\times 10^{7}$   \\
\hline
$B_c^+ \rightarrow B^{+} K_{S}^0\rightarrow\bar{D}^0 \rho^{+} K_{S}^0\rightarrow  K^+ e^-\bar{\nu}_e  \pi^{+} \pi^0 \pi^+\pi^- $ & $\left(1.14\sim1.60\right)\times 10^{7}$   \\
\hline
$B_c^+ \rightarrow B^{+} K_{S}^0\rightarrow\bar{D}^0 \rho^{+} K_{S}^0\rightarrow  K^+ \pi^-  \pi^{+} \pi^0 \pi^+\pi^- $ & $\left(1.03\sim1.44\right)\times 10^{7}$   \\
\hline
$B_c^+ \rightarrow B^{+} K_{S}^0\rightarrow\bar{D}^0 \rho^{+} K_{S}^0\rightarrow  K^+ \pi^- \pi^0  \pi^{+} \pi^0 \pi^+\pi^- $ & $\left(2.80\sim3.96\right)\times 10^{6}$   \\
\hline
$B_c^+ \rightarrow B^{+} K_{S}^0\rightarrow\bar{D}^* (2007)^0 \pi^{+} K_{S}^0\rightarrow\bar{D}^0 \pi^0 \pi^{+} K_{S}^0 \rightarrow K^+  \pi^- \pi^0\pi^0 \pi^{+} \pi^+\pi^-  $ & $\left(1.25\sim1.56\right)\times 10^{7}$   \\
\hline
$B_c^+ \rightarrow B^{+} K_{S}^0\rightarrow\bar{D}^* (2007)^0 \pi^{+}\pi^+\pi^- K_{S}^0\rightarrow\bar{D}^0 \pi^0 \pi^{+}\pi^+\pi^- K_{S}^0 \rightarrow K^+  \pi^- \pi^0 \pi^0 \pi^{+} \pi^+\pi^-  \pi^+\pi^-  $ & $\left(5.70\sim7.84\right)\times 10^{6}$   \\
\hline
$B_c^+ \rightarrow B^{+} K_{S}^0\rightarrow J/\psi K^{+}K_{S}^0\rightarrow e^{+} e^- K^+   \pi^+\pi^-  $ & $\left(0.95\sim1.18\right)\times 10^{8}$   \\
\hline
$B_c^+ \rightarrow B^{+} K_{L}^0\rightarrow\bar{D}^0 l^{+}\nu_l K_{L}^0\rightarrow K^+ e^-\bar{\nu}_e l^{+}\nu_l \pi^\pm e^\mp \nu_e $ & $\left(0.95\sim1.17\right)\times 10^{7}$   \\
\hline
$B_c^+ \rightarrow B^{+} K_{L}^0\rightarrow\bar{D}^0 l^{+}\nu_l K_{L}^0\rightarrow K^+ \pi^- l^{+}\nu_l \pi^\pm e^\mp \nu_e $ & $\left(0.85\sim1.05\right)\times 10^{7}$   \\
\hline
$B_c^+ \rightarrow B^{+} K_{L}^0\rightarrow\bar{D}^0 l^{+}\nu_l K_{L}^0\rightarrow K^+ \pi^- \pi^0 l^{+}\nu_l \pi^\pm e^\mp \nu_e $ & $\left(2.32\sim2.91\right)\times 10^{6}$   \\
\hline
$B_c^+ \rightarrow B^{+} K_{L}^0\rightarrow\bar{D}^0 l^{+}\nu_l K_{L}^0\rightarrow K^+ \pi^- \pi^0 l^{+}\nu_l \pi^+ \pi^- \pi^0 $ & $\left(7.49\sim9.40\right)\times 10^{6}$   \\
\hline
$B_c^+ \rightarrow B^{+} K_{L}^0\rightarrow\bar{D}^* (2007)^0 l^{+}\nu_l K_{L}^0\rightarrow\bar{D}^0 \pi^0 l^{+}\nu_l K_{L}^0 \rightarrow K^+  e^-\bar{\nu}_e \pi^0 l^{+}\nu_l \pi^\pm e^\mp \nu_e  $ & $\left(6.06\sim7.55 \right)\times 10^{6}$   \\
\hline
$B_c^+ \rightarrow B^{+} K_{L}^0\rightarrow\bar{D}^* (2007)^0 l^{+}\nu_l K_{L}^0\rightarrow\bar{D}^0 \pi^0 l^{+}\nu_l K_{L}^0 \rightarrow K^+  \pi^- \pi^0 l^{+}\nu_l \pi^\pm e^\mp \nu_e  $ & $\left(5.44\sim6.77\right)\times 10^{6}$   \\
\hline
$B_c^+ \rightarrow B^{+} K_{L}^0\rightarrow\bar{D}^* (2007)^0 l^{+}\nu_l K_{L}^0\rightarrow\bar{D}^0 \pi^0 l^{+}\nu_l K_{L}^0 \rightarrow K^+  \pi^- \pi^0 \pi^0 l^{+}\nu_l \pi^\pm e^\mp \nu_e  $ & $\left(1.49\sim1.87\right)\times 10^{6}$   \\
\hline
$B_c^+ \rightarrow B^{+} K_{L}^0\rightarrow\bar{D}^* (2007)^0 l^{+}\nu_l K_{L}^0\rightarrow\bar{D}^0 \pi^0 l^{+}\nu_l K_{L}^0 \rightarrow K^+  \pi^- \pi^0 \pi^0 l^{+}\nu_l \pi^+ \pi^- \pi^0  $ & $\left(4.80\sim6.04\right)\times 10^{6}$   \\
\hline
$B_c^+ \rightarrow B^{+} K_{L}^0\rightarrow\bar{D}^0 \pi^{+} K_{L}^0\rightarrow K^+ \pi^- \pi^0 \pi^{+}\pi^\pm e^\mp \nu_e$ & $\left(1.17\sim1.45\right)\times 10^{7}$   \\
\hline
$B_c^+ \rightarrow B^{+} K_{L}^0\rightarrow\bar{D}^0 \rho^{+} K_{L}^0\rightarrow  K^+ \pi^-  \pi^{+} \pi^0 \pi^\pm e^\mp \nu_e $ & $\left(1.41\sim1.98\right)\times 10^{7}$   \\
\hline
$B_c^+ \rightarrow B^{+} K_{L}^0\rightarrow\bar{D}^0 \rho^{+} K_{L}^0\rightarrow  K^+ \pi^- \pi^0 \pi^{+} \pi^0 \pi^\pm e^\mp \nu_e $ & $\left(3.86\sim5.46\right)\times 10^{6}$   \\
\hline
$B_c^+ \rightarrow B^{+} K_{L}^0\rightarrow\bar{D}^0 \rho^{+} K_{L}^0\rightarrow  K^+ \pi^- \pi^0 \pi^{+} \pi^0 \pi^+ \pi^- \pi^0 $ & $\left(1.25\sim1.76\right)\times 10^{7}$   \\
\hline
$B_c^+ \rightarrow B^{+} K_{L}^0\rightarrow\bar{D}^* (2007)^0 \pi^{+}\pi^+\pi^- K_{L}^0\rightarrow\bar{D}^0 \pi^0 \pi^{+}\pi^+\pi^- K_{L}^0 \rightarrow K^+  \pi^- \pi^0 \pi^0 \pi^{+} \pi^+\pi^-  \pi^\pm e^\mp \nu_e  $ & $\left(0.79\sim1.08\right)\times 10^{7}$   \\
\hline
$B_c^+ \rightarrow B^{+} K_{L}^0\rightarrow J/\psi K^{+}K_{L}^0\rightarrow e^{+} e^- K^+   \pi^\pm e^\mp \nu_e  $ & $\left(1.32\sim1.62\right)\times 10^{8}$   \\
\hline
\end{tabular}
\end{center}
\end{table}

In the same way, the numbers of $B_c^\pm$ events-times-efficiency needed for observing the $K_S^0-K_L^0$ asymmetries at three standard deviations (3$\sigma$) level are
\begin{align}
(\epsilon_f N)_{K_{S,L}^0}^+=\frac{9}{\left[{\mathcal B}(B_c^+ \rightarrow B^{+} K_{S}^0\rightarrow f_{B^+} K_{S}^0)+{\mathcal B}(B_c^+ \rightarrow B^{+} K_{L}^0\rightarrow f_{B^+} K_{L}^0)\right]\cdot \left|R(B_c^+\rightarrow B^+ K_{S,L}^0)\right|},\label{Eq:numneedkslbplus}
\end{align}
and
\begin{align}
(\epsilon_f N)_{K_{S,L}^0}^-=\frac{9}{\left[{\mathcal B}(B_c^- \rightarrow B^{-} K_{S}^0\rightarrow f_{B^-} K_{S}^0)+{\mathcal B}(B_c^- \rightarrow B^{-} K_{L}^0\rightarrow f_{B^-} K_{L}^0)\right]\cdot \left|R(B_c^-\rightarrow B^- K_{S,L}^0)\right|}.\label{Eq:numneedkslbminus}
\end{align}
From Particle Data Group~\cite{Zyla:2020zbs}, Table~\ref{totbranchratio}, Eq.(\ref{Eq:numksklasy1}) and Eq.(\ref{Eq:numksklasy2}), we can calculate the numerical results of $(\epsilon_f N)_{K_{S,L}^0}^+$ and $(\epsilon_f N)_{K_{S,L}^0}^-$, which are listed in Table~\ref{tabnumksyasobe}.
\begin{table}[t]
\begin{center}
\caption{\label{tabnumksyasobe} \small The numerical results of $(\epsilon_f N)_{K_{S,L}^0}^\pm$  for the $B_c^{\pm}\rightarrow B^{\pm} K_{S,L}^0\rightarrow f_{B^\pm} K_{S,L}^0$ decays.}
\vspace{0.1cm}
\doublerulesep 0.8pt \tabcolsep 0.18in
\scriptsize
\begin{tabular}{c|c}
\hline
the decay channel & $(\epsilon_f N)_{K_{S,L}^0}^+$ ($ (\epsilon_f N)_{K_{S,L}^0}^-$) \\
\hline
$B_c^+ \rightarrow B^{+} K_{S,L}^0\rightarrow\bar{D}^0 l^{+}\nu_l K_{S,L}^0\rightarrow K^+ e^-\bar{\nu}_e l^{+}\nu_l K_{S,L}^0 $ & $\left(1.99\sim2.35\right)\times 10^{7}$ \\
\hline
$B_c^- \rightarrow B^{-} K_{S,L}^0\rightarrow D^0 l^{-} \bar{\nu}_l K_{S,L}^0\rightarrow K^- e^+ \nu_e l^{-}\bar{\nu}_l K_{S,L}^0 $ &  $\left(2.11\sim2.49\right)\times 10^{7}$\\
\hline
$B_c^+ \rightarrow B^{+} K_{S,L}^0\rightarrow\bar{D}^0 l^{+}\nu_l K_{S,L}^0\rightarrow K^+ \pi^- l^{+}\nu_l K_{S,L}^0$ & $\left(1.79\sim2.11\right)\times10^{7}$ \\
\hline
$B_c^- \rightarrow B^{-} K_{S,L}^0\rightarrow D^0 l^{-}\bar{\nu}_l K_{S,L}^0\rightarrow K^- \pi^+ l^{-}\bar{\nu}_l K_{S,L}^0$ & $\left(1.89\sim2.23\right)\times10^{7}$   \\
\hline
$B_c^+ \rightarrow B^{+} K_{S,L}^0\rightarrow\bar{D}^0 l^{+}\nu_l K_{S,L}^0\rightarrow K^+ \pi^- \pi^0 l^{+}\nu_l K_{S,L}^0 $ & $\left(4.88\sim 5.82\right)\times 10^{6}$    \\
\hline
$B_c^- \rightarrow B^{-} K_{S,L}^0\rightarrow D^0 l^{-}\bar{\nu}_l K_{S,L}^0\rightarrow K^- \pi^+\pi^0 l^{-}\bar{\nu}_l K_{S,L}^0$ & $\left(5.16\sim 6.16\right)\times 10^{6}$   \\
\hline
$B_c^+ \rightarrow B^{+} K_{S,L}^0\rightarrow\bar{D}^* (2007)^0 l^{+}\nu_l K_{S,L}^0\rightarrow\bar{D}^0 \pi^0 l^{+}\nu_l K_{S,L}^0 \rightarrow K^+  e^-\bar{\nu}_e \pi^0 l^{+}\nu_l K_{S,L}^0 $ & $\left(1.28\sim 1.51\right)\times 10^{7}$   \\
\hline
$B_c^- \rightarrow B^{-} K_{S,L}^0\rightarrow D^* (2007)^0 l^{-} \bar{\nu}_l K_{S,L}^0\rightarrow D^0 \pi^0 l^{-}\bar{\nu}_l K_{S,L}^0 \rightarrow K^-  e^+ \nu_e \pi^0 l^{-}\bar{\nu}_l K_{S,L}^0 $ & $\left(1.35\sim 1.60\right)\times 10^{7}$  \\
\hline
$B_c^+ \rightarrow B^{+} K_{S,L}^0\rightarrow\bar{D}^* (2007)^0 l^{+}\nu_l K_{S,L}^0\rightarrow\bar{D}^0 \pi^0 l^{+}\nu_l K_{S,L}^0 \rightarrow K^+  \pi^- \pi^0 l^{+}\nu_l K_{S,L}^0 $ & $\left(1.15\sim1.35\right)\times 10^{7}$  \\
\hline
$B_c^- \rightarrow B^{-} K_{S,L}^0\rightarrow D^* (2007)^0 l^{-} \bar{\nu}_l K_{S,L}^0\rightarrow D^0 \pi^0 l^{-}\bar{\nu}_l K_{S,L}^0 \rightarrow K^-  \pi^+ \pi^0 l^{-}\bar{\nu}_l K_{S,L}^0 $ & $\left(1.21\sim1.43\right)\times 10^{7}$ \\
\hline
$B_c^+ \rightarrow B^{+} K_{S,L}^0\rightarrow\bar{D}^* (2007)^0 l^{+}\nu_l K_{S,L}^0\rightarrow\bar{D}^0 \pi^0 l^{+}\nu_l K_{S,L}^0 \rightarrow K^+  \pi^- \pi^0\pi^0 l^{+}\nu_l K_{S,L}^0 $ & $\left(3.13\sim3.74\right)\times 10^{6}$  \\
\hline
$B_c^- \rightarrow B^{-} K_{S,L}^0\rightarrow D^* (2007)^0 l^{-} \bar{\nu}_l K_{S,L}^0\rightarrow D^0 \pi^0 l^{-}\bar{\nu}_l K_{S,L}^0 \rightarrow K^-  \pi^+ \pi^0\pi^0 l^{-}\bar{\nu}_l K_{S,L}^0 $ & $\left(3.31\sim3.96\right)\times 10^{6}$  \\
\hline
$B_c^+ \rightarrow B^{+} K_{S,L}^0\rightarrow\bar{D}^0 \pi^{+} K_{S,L}^0\rightarrow K^+ \pi^- \pi^0 \pi^{+}K_{S,L}^0 $ & $\left(2.46\sim 2.91\right)\times 10^{7}$  \\
\hline
$B_c^- \rightarrow B^{-} K_{S,L}^0\rightarrow D^0 \pi^{-} K_{S,L}^0\rightarrow K^- \pi^+ \pi^0 \pi^{-} K_{S,L}^0 $ & $\left(2.60\sim 3.08\right)\times 10^{7}$  \\
\hline
$B_c^+ \rightarrow B^{+} K_{S,L}^0\rightarrow\bar{D}^0 \rho^{+} K_{S,L}^0\rightarrow  K^+ e^-\bar{\nu}_e  \pi^{+} \pi^0 K_{S,L}^0 $ & $\left(3.29\sim4.47\right)\times 10^{7}$    \\
\hline
$B_c^- \rightarrow B^{-} K_{S,L}^0\rightarrow D^0 \rho^{-} K_{S,L}^0\rightarrow  K^- e^+ \nu_e  \pi^{-} \pi^0 K_{S,L}^0 $ & $\left(3.48\sim4.73\right)\times 10^{7}$   \\
\hline
$B_c^+ \rightarrow B^{+} K_{S,L}^0\rightarrow\bar{D}^0 \rho^{+} K_{S,L}^0\rightarrow  K^+ \pi^-  \pi^{+} \pi^0 K_{S,L}^0 $ & $\left(2.95\sim4.01\right)\times 10^{7}$   \\
\hline
$B_c^- \rightarrow B^{-} K_{S,L}^0\rightarrow D^0 \rho^{-} K_{S,L}^0\rightarrow  K^- \pi^+  \pi^{-} \pi^0 K_{S,L}^0 $ & $\left(3.12\sim4.24\right)\times 10^{7}$   \\
\hline
$B_c^+ \rightarrow B^{+} K_{S,L}^0\rightarrow\bar{D}^0 \rho^{+} K_{S,L}^0\rightarrow  K^+ \pi^- \pi^0  \pi^{+} \pi^0 K_{S,L}^0 $ & $\left(0.81\sim1.10\right)\times 10^{7}$  \\
\hline
$B_c^- \rightarrow B^{-} K_{S,L}^0\rightarrow D^0 \rho^{-} K_{S,L}^0\rightarrow  K^- \pi^+ \pi^0  \pi^{-} \pi^0 K_{S,L}^0 $ & $\left(0.85\sim1.17\right)\times 10^{7}$  \\
\hline
$B_c^+ \rightarrow B^{+} K_{S,L}^0\rightarrow\bar{D}^* (2007)^0 \pi^{+} K_{S,L}^0\rightarrow\bar{D}^0 \pi^0 \pi^{+} K_{S,L}^0 \rightarrow K^+  \pi^- \pi^0\pi^0 \pi^{+} K_{S,L}^0  $ &$\left(3.62\sim4.32\right)\times 10^{7}$  \\
\hline
$B_c^- \rightarrow B^{-} K_{S,L}^0\rightarrow D^* (2007)^0 \pi^{-} K_{S,L}^0\rightarrow D^0 \pi^0 \pi^{-} K_{S,L}^0 \rightarrow K^-  \pi^+ \pi^0\pi^0 \pi^{-} K_{S,L}^0  $ &  $\left(3.83\sim4.57\right)\times 10^{7}$   \\
\hline
$B_c^+ \rightarrow B^{+} K_{S,L}^0\rightarrow\bar{D}^* (2007)^0 \pi^{+}\pi^+\pi^- K_{S,L}^0\rightarrow\bar{D}^0 \pi^0 \pi^{+}\pi^+\pi^- K_{S,L}^0 \rightarrow K^+  \pi^- \pi^0 \pi^0 \pi^{+} \pi^+\pi^-  K_{S,L}^0  $ & $\left(1.64\sim2.18\right)\times 10^{7}$ \\
\hline
$B_c^- \rightarrow B^{-} K_{S,L}^0\rightarrow D^* (2007)^0 \pi^{-}\pi^-\pi^+ K_{S,L}^0\rightarrow D^0 \pi^0 \pi^{-}\pi^-\pi^+ K_{S,L}^0 \rightarrow K^-  \pi^+ \pi^0 \pi^0 \pi^{-}\pi^-\pi^+  K_{S,L}^0  $ & $\left(1.74\sim2.31\right)\times 10^{7}$  \\
\hline
$B_c^+ \rightarrow B^{+} K_{S,L}^0\rightarrow J/\psi K^{+} K_{S,L}^0\rightarrow e^{+} e^- K^+   K_{S,L}^0  $ & $\left(2.78\sim3.24\right)\times 10^{8}$ \\
\hline
$B_c^- \rightarrow B^{-} K_{S,L}^0\rightarrow J/\psi K^{-} K_{S,L}^0\rightarrow e^{+} e^- K^-   K_{S,L}^0  $ & $\left(2.94\sim3.43\right)\times 10^{8}$ \\
\hline
\end{tabular}
\end{center}
\end{table}

Again, the numbers of $B_c^\pm$ events-times-efficiency needed for testing CP violation with 3$\sigma$ significance can be derived as follows
\begin{align}
(\epsilon_f N)_{CP}=&\frac{9}{\left[{\mathcal B}(B_c^+ \rightarrow B^{+} K_{S,L}^0\rightarrow f_{B^+} f_{K_{S,L}^0})+{\mathcal B}(B_c^- \rightarrow B^{-} K_{S,L}^0\rightarrow f_{B^-} f_{K_{S,L}^0})\right]\cdot \left| {\mathcal A}_{CP}^{K_{S,L}^{0}}\right|}\nonumber\\
\approx &\frac{9}{2\cdot {\mathcal B}(B_c^+ \rightarrow B^{+} K_{S,L}^0\rightarrow f_{B^+} f_{K_{S,L}^0})\cdot \left|{\mathcal A}_{CP}^{K_{S,L}^{0}}\right|}.\label{Eq:numneedcpv}
\end{align}
As discussed in section~\ref{sec:cpvcptvkskl} and section~\ref{sec:numcalution}, the CP asymmetries can be observed only in the $B_c^\pm \rightarrow B^{\pm} K_{S,L}^0$ decays and the semileptonic decays of the $B^\pm$ to charmed meson, so we only calculate the numbers of $B_c^\pm$ events-times-efficiency needed in these decays,which are shown in Table~\ref{tabnumcpvio}. Meanwhile, we note that the decays $B_c^{\pm}\rightarrow B^{\pm} K_{S}^0\rightarrow f_{B^\pm}\pi^+\pi^-$ are more favorable than the decays $B_c^{\pm}\rightarrow B^{\pm} K_{L}^0\rightarrow f_{B^\pm}\pi^\pm e^\mp \nu_e (\pi^+ \pi^- \pi^0)$ to observe in experiment,  because the number of the final state particles in $K_{S}^0\rightarrow \pi^+\pi^-$ decay is less than that in $K_{L}^0\rightarrow \pi^\pm e^\mp \nu_e (\pi^+ \pi^- \pi^0)$ decays.
\begin{table}[t]
\begin{center}
\caption{\label{tabnumcpvio} \small The numerical results of $(\epsilon_f N)_{CP}$  for the $B_c^{\pm}\rightarrow B^{\pm} K_{S,L}^0\rightarrow f_{B^\pm} f_{K_{S,L}^0}$ decays.}
\vspace{0.1cm}
\doublerulesep 0.8pt \tabcolsep 0.18in
\scriptsize
\begin{tabular}{c|c}
\hline
the decay channel & $(\epsilon_f N)_{CP}$ \\
\hline
$B_c^+ \rightarrow B^{+} K_{S}^0\rightarrow\bar{D}^0 l^{+}\nu_l K_{S}^0\rightarrow K^+ e^-\bar{\nu}_e l^{+}\nu_l \pi^+\pi^- $ & $\left(0.85\sim1.05\right)\times 10^{9}$   \\
\hline
$B_c^+ \rightarrow B^{+} K_{S}^0\rightarrow\bar{D}^0 l^{+}\nu_l K_{S}^0\rightarrow K^+ \pi^- l^{+}\nu_l \pi^+\pi^- $ & $\left(7.60\sim9.44\right)\times10^{8}$   \\
\hline
$B_c^+ \rightarrow B^{+} K_{S}^0\rightarrow\bar{D}^0 l^{+}\nu_l K_{S}^0\rightarrow K^+ \pi^- \pi^0 l^{+}\nu_l \pi^+\pi^- $ & $\left(2.08\sim 2.60\right)\times 10^{8}$   \\
\hline
$B_c^+ \rightarrow B^{+} K_{S}^0\rightarrow\bar{D}^* (2007)^0 l^{+}\nu_l K_{S}^0\rightarrow\bar{D}^0 \pi^0 l^{+}\nu_l K_{S}^0 \rightarrow K^+  e^-\bar{\nu}_e \pi^0 l^{+}\nu_l \pi^+\pi^- $ & $\left(5.43\sim 6.76\right)\times 10^{8}$   \\
\hline
$B_c^+ \rightarrow B^{+} K_{S}^0\rightarrow\bar{D}^* (2007)^0 l^{+}\nu_l K_{S}^0\rightarrow\bar{D}^0 \pi^0 l^{+}\nu_l K_{S}^0 \rightarrow K^+  \pi^- \pi^0 l^{+}\nu_l \pi^+\pi^- $ & $\left(4.87\sim6.06\right)\times 10^{8}$   \\
\hline
$B_c^+ \rightarrow B^{+} K_{S}^0\rightarrow\bar{D}^* (2007)^0 l^{+}\nu_l K_{S}^0\rightarrow\bar{D}^0 \pi^0 l^{+}\nu_l K_{S}^0 \rightarrow K^+  \pi^- \pi^0\pi^0 l^{+}\nu_l \pi^+\pi^-  $ & $\left(1.33\sim1.67\right)\times 10^{8}$   \\
\hline
$B_c^+ \rightarrow B^{+} K_{L}^0\rightarrow\bar{D}^0 l^{+}\nu_l K_{L}^0\rightarrow K^+ e^-\bar{\nu}_e l^{+}\nu_l \pi^\pm e^\mp \nu_e $ & $\left(1.59\sim1.97\right)\times 10^{9}$   \\
\hline
$B_c^+ \rightarrow B^{+} K_{L}^0\rightarrow\bar{D}^0 l^{+}\nu_l K_{L}^0\rightarrow K^+ \pi^- l^{+}\nu_l \pi^\pm e^\mp \nu_e $ & $\left(1.42\sim1.77\right)\times 10^{9}$   \\
\hline
$B_c^+ \rightarrow B^{+} K_{L}^0\rightarrow\bar{D}^0 l^{+}\nu_l K_{L}^0\rightarrow K^+ \pi^- \pi^0 l^{+}\nu_l \pi^\pm e^\mp \nu_e $ & $\left(3.88\sim4.88\right)\times 10^{8}$   \\
\hline
$B_c^+ \rightarrow B^{+} K_{L}^0\rightarrow\bar{D}^0 l^{+}\nu_l K_{L}^0\rightarrow K^+ \pi^- \pi^0 l^{+}\nu_l \pi^+ \pi^- \pi^0 $ & $\left(1.26\sim1.58\right)\times 10^{9}$   \\
\hline
$B_c^+ \rightarrow B^{+} K_{L}^0\rightarrow\bar{D}^* (2007)^0 l^{+}\nu_l K_{L}^0\rightarrow\bar{D}^0 \pi^0 l^{+}\nu_l K_{L}^0 \rightarrow K^+  e^-\bar{\nu}_e \pi^0 l^{+}\nu_l \pi^\pm e^\mp \nu_e  $ & $\left(1.02\sim1.27 \right)\times 10^{9}$   \\
\hline
$B_c^+ \rightarrow B^{+} K_{L}^0\rightarrow\bar{D}^* (2007)^0 l^{+}\nu_l K_{L}^0\rightarrow\bar{D}^0 \pi^0 l^{+}\nu_l K_{L}^0 \rightarrow K^+  \pi^- \pi^0 l^{+}\nu_l \pi^\pm e^\mp \nu_e  $ & $\left(0.91\sim1.14\right)\times 10^{9}$   \\
\hline
$B_c^+ \rightarrow B^{+} K_{L}^0\rightarrow\bar{D}^* (2007)^0 l^{+}\nu_l K_{L}^0\rightarrow\bar{D}^0 \pi^0 l^{+}\nu_l K_{L}^0 \rightarrow K^+  \pi^- \pi^0 \pi^0 l^{+}\nu_l \pi^\pm e^\mp \nu_e  $ & $\left(2.49\sim3.13\right)\times 10^{8}$   \\
\hline
$B_c^+ \rightarrow B^{+} K_{L}^0\rightarrow\bar{D}^* (2007)^0 l^{+}\nu_l K_{L}^0\rightarrow\bar{D}^0 \pi^0 l^{+}\nu_l K_{L}^0 \rightarrow K^+  \pi^- \pi^0 \pi^0 l^{+}\nu_l \pi^+ \pi^- \pi^0  $ & $\left(0.81\sim1.01\right)\times 10^{9}$   \\
\hline
\end{tabular}
\end{center}
\end{table}

From Table~\ref{tabnumbrobe}, Table~\ref{tabnumksyasobe} and Table~\ref{tabnumcpvio}, we can see that the numbers of the $B_c^{\pm}$ events-times-efficiency, which are needed to observe the branching ratios, $K_S^0-K_L^0$ asymmetries and CP asymmetries, are of the order of $10^6 \sim 10^9$ in the listed processes, so they are possible to be observed at the LHC experiment and the HL-LHC experiment.
\section{Conclusions}
\label{sec:conclusions}
In conclusion, the large $B_c^{\pm}$ sample, which can be produced at the LHC experiment and the HL-LHC experiment, will make it an ideal place to study the $B_c^{\pm}$ decays and investigate the related physics. We study the effect of $K^0-\bar{K}^0$ mixing on the branching ratios, CP asymmetries, $K_S^0-K_L^0$ asymmetries and CPT violations in the $B_c^{\pm}\rightarrow B^{\pm} K_{S,L}^0$ decays. We find that some of the $B_c^{\pm}\rightarrow B^{\pm} K_{S,L}^0\rightarrow f_{B^{\pm}} f_{K_{S,L}^0}$ decay chains have large branching ratios, whose maximum value can reach $7.48\times 10^{-6}$. The CP asymmetries in the $B_c^{\pm}\rightarrow B^{\pm} K_{S,L}^0$ decays are dominated by $K^0-\bar{K}^0$ mixing and can exceed the order of $10^{-3}$. The most promising processes to observe the CP violation are $B_c^+ \rightarrow B^{+} K_{S}^0\rightarrow\bar{D}^0 l^{+}\nu_l K_{S}^0\rightarrow K^+ e^-\bar{\nu}_e l^{+}\nu_l \pi^+\pi^-$, $B_c^+ \rightarrow B^{+} K_{S}^0\rightarrow\bar{D}^0 l^{+}\nu_l K_{S}^0\rightarrow K^+ \pi^- l^{+}\nu_l \pi^+\pi^-$, $B_c^+ \rightarrow B^{+} K_{S}^0\rightarrow\bar{D}^0 l^{+}\nu_l K_{S}^0\rightarrow K^+ \pi^- \pi^0 l^{+}\nu_l \pi^+\pi^-$, $B_c^+ \rightarrow B^{+} K_{S}^0\rightarrow\bar{D}^* (2007)^0 l^{+}\nu_l K_{S}^0\rightarrow\bar{D}^0 \pi^0 l^{+}\nu_l K_{S}^0 \rightarrow K^+  e^-\bar{\nu}_e \pi^0 l^{+}\nu_l \pi^+\pi^-$, $B_c^+ \rightarrow B^{+} K_{S}^0\rightarrow\bar{D}^* (2007)^0 l^{+}\nu_l K_{S}^0\rightarrow\bar{D}^0 \pi^0 l^{+}\nu_l K_{S}^0 \rightarrow K^+  \pi^- \pi^0 l^{+}\nu_l \pi^+\pi^-$ and $B_c^+ \rightarrow B^{+} K_{S}^0\rightarrow\bar{D}^* (2007)^0 l^{+}\nu_l K_{S}^0$ $\rightarrow\bar{D}^0 \pi^0 l^{+}\nu_l K_{S}^0 \rightarrow K^+  \pi^- \pi^0\pi^0 l^{+}\nu_l \pi^+\pi^-$, for which about $1.05\times 10^9$, $9.44\times 10^8$, $2.60\times 10^8$, $6.76\times 10^8$, $6.06\times 10^8$ and $1.67\times 10^8$ $B_c^{\pm}$ events-times-efficiency are needed respectively, if CP violation is observed at 3$\sigma$ level.

The $K_S^0-K_L^0$ asymmetries in the $B_c^{\pm}\rightarrow B^{\pm} K_{S,L}^0$ decays are also studied. The numerical results of the $K_S^0-K_L^0$ asymmetries can reach as large as $0.11$. Together with the branching ratios for the $B_c^{\pm} \rightarrow B^{\pm} K_{S,L}^0\rightarrow f_{B^{\pm}} K_{S,L}^0$ decay chains and the $K_S^0-K_L^0$ asymmetries, we calculate the numbers of the $B_c^{\pm}$ events-times-efficiency needed to establish the $K_S^0-K_L^0$ asymmetries to 3 standard deviations in the decay processes with a large branching ratios. The range of the numbers of the $B_c^{\pm}$ events-times-efficiency needed to observe $K_S^0-K_L^0$ asymmetries at a significance of 3$\sigma$ in these decays is from  $3.74\times 10^6$ to $3.43\times 10^8$, which is measurable at the LHC experiment. We investigate the possibility of extracting the CPT violation parameter $z$ in the $B_c^{\pm} \rightarrow B^{\pm} K_{S,L}^0\rightarrow f_{B^{\pm}} f_{K_{S,L}^0}$ decays. We find that the decays $B_c^+ \rightarrow B^{+} K_{S,L}^0\rightarrow\bar{D}^0 l^{+}\nu_l K_{S,L}^0\rightarrow K^+ e^-\bar{\nu}_e l^{+}\nu_l K_{S,L}^0$ and $B_c^+ \rightarrow B^{+} K_{S,L}^0\rightarrow\bar{D}^* (2007)^0 l^{+}\nu_l K_{S,L}^0\rightarrow\bar{D}^0 \pi^0 l^{+}\nu_l K_{S,L}^0 \rightarrow K^+  e^-\bar{\nu}_e \pi^0 l^{+}\nu_l K_{S,L}^0$ can in principle constraint the parameter $z$ if the measurements of the parameters $m_L-m_S$, $\Gamma_L$, $\Gamma_S$, $p$ and $q$ have high precision and the values of $t_0$ and $t_1$ are consistent with the event selection criteria in experiment.
\section*{Acknowledgements}
The work was supported by the National Natural Science Foundation of China (Contract Nos. 11675137, 11805077) and the Key Scientific Research Projects of Colleges and Universities in Henan Province (Contract No. 18A140029).
\begin{appendix}
\numberwithin{equation}{section}
\section{The relations used in ${\mathcal A}_{CP}^{K_{S,L}^{0}}$  calculations }
\label{sec:appendixcpcal}
Below we present the relations used in evaluating the expression of ${\mathcal A}_{CP}^{K_{S,L}^{0}}$ in section~\ref{sec:cpvcptvkskl}
\begin{align}
&\Gamma(B_c^-\rightarrow B^{-}K_{S,L}^0)\cdot\Gamma(B^{-}\rightarrow f_{B^-})-\Gamma(B_c^+\rightarrow B^{+}K_{S,L}^0)\cdot\Gamma(B^{+}\rightarrow \bar{f}_{B^-})\nonumber\\
&=\Gamma(B_c^+\rightarrow B^{+}K_{S,L}^0)\cdot\left(\Gamma(B^{-}\rightarrow f_{B^-})-\Gamma(B^{+}\rightarrow \bar{f}_{B^-})\right)\nonumber\\
&~~~~~~~~~~~~~~~~+\left(\Gamma(B_c^-\rightarrow B^{-}K_{S,L}^0)-\Gamma(B_c^+\rightarrow B^{+}K_{S,L}^0)\right)\cdot \Gamma(B^{-}\rightarrow f_{B^-})\nonumber\\
&=\Gamma(B_c^-\rightarrow B^{-}K_{S,L}^0)\cdot\left(\Gamma(B^{-}\rightarrow f_{B^-})-\Gamma(B^{+}\rightarrow \bar{f}_{B^-})\right)\nonumber\\
&~~~~~~~~~~~~~~~~+\left(\Gamma(B_c^-\rightarrow B^{-}K_{S,L}^0)-\Gamma(B_c^+\rightarrow B^{+}K_{S,L}^0)\right)\cdot \Gamma(B^{+}\rightarrow \bar{f}_{B^-}),\label{Eq:bianhuanrelation1}
\end{align}
and
\begin{align}
&\Gamma(B_c^-\rightarrow B^{-}K_{S,L}^0)\cdot\Gamma(B^{-}\rightarrow f_{B^-})+\Gamma(B_c^+\rightarrow B^{+}K_{S,L}^0)\cdot\Gamma(B^{+}\rightarrow \bar{f}_{B^-})\nonumber\\
&=\Gamma(B_c^+\rightarrow B^{+}K_{S,L}^0)\cdot\left(\Gamma(B^{-}\rightarrow f_{B^-})+\Gamma(B^{+}\rightarrow \bar{f}_{B^-})\right)\nonumber\\
&~~~~~~~~~~~~~~~~+\left(\Gamma(B_c^-\rightarrow B^{-}K_{S,L}^0)-\Gamma(B_c^+\rightarrow B^{+}K_{S,L}^0)\right)\cdot \Gamma(B^{-}\rightarrow f_{B^-})\nonumber\\
&=\left(\Gamma(B_c^-\rightarrow B^{-}K_{S,L}^0)+\Gamma(B_c^+\rightarrow B^{+}K_{S,L}^0)\right)\cdot \Gamma(B^{-}\rightarrow f_{B^-})\nonumber\\
&~~~~~~~~~~~~~~~~-\Gamma(B_c^+\rightarrow B^{+}K_{S,L}^0)\cdot\left(\Gamma(B^{-}\rightarrow f_{B^-})-\Gamma(B^{+}\rightarrow \bar{f}_{B^-})\right).\label{Eq:bianhuanrelation2}
\end{align}
\section{The ${\mathcal A}_{CPT}$  calculations}
\label{sec:appendixcptcal}
To calculate the expression of ${\mathcal A}_{CPT}$, the following relations are employed
\begin{align}
&\left[\Gamma(B_c^{-}\rightarrow  B^- K_S^0)\Gamma(B^{-}\rightarrow f_{B^-})- \Gamma(B_c^{+}\rightarrow  B^+ K_S^0)\Gamma(B^{+}\rightarrow \bar{f}_{B^-})\right]\nonumber\\
&-\left[\Gamma(B_c^{-}\rightarrow  B^- K_L^0)\Gamma(B^{-}\rightarrow f_{B^-})- \Gamma(B_c^{+}\rightarrow  B^+ K_L^0)\Gamma(B^{+}\rightarrow \bar{f}_{B^-})\right]\nonumber\\
&=\left[\Gamma(B_c^{+}\rightarrow  B^+ K_S^0)-\Gamma(B_c^{+}\rightarrow  B^+ K_L^0)\right]\cdot\left[\Gamma(B^{-}\rightarrow f_{B^-}) -\Gamma(B^{+}\rightarrow \bar{f}_{B^-})\right]+\left[\left(\Gamma\left(B_c^{-}\right.\right.\right.\nonumber\\
&\left.\left.\left.\rightarrow  B^- K_S^0\right)-\Gamma(B_c^{+}\rightarrow  B^+ K_S^0)\right)-\left(\Gamma(B_c^{-}\rightarrow  B^- K_L^0)-\Gamma(B_c^{+}\rightarrow  B^+ K_L^0)\right)\right]\cdot \Gamma(B^{-}\rightarrow f_{B^-}),\label{Eq:cptasyobserre1}
\end{align}
and
\begin{align}
&\Gamma(B_c^{-}\rightarrow  B^- K_S^0)\Gamma(B^{-}\rightarrow f_{B^-})+ \Gamma(B_c^{+}\rightarrow  B^+ K_S^0)\Gamma(B^{+}\rightarrow \bar{f}_{B^-})\nonumber\\
&+\Gamma(B_c^{-}\rightarrow  B^- K_L^0)\Gamma(B^{-}\rightarrow f_{B^-})+ \Gamma(B_c^{+}\rightarrow  B^+ K_L^0)\Gamma(B^{+}\rightarrow \bar{f}_{B^-})\nonumber\\
&=\left[\Gamma(B_c^{+}\rightarrow  B^+ K_S^0)+\Gamma(B_c^{+}\rightarrow  B^+ K_L^0)\right]\cdot\left[\Gamma(B^{-}\rightarrow f_{B^-})+\Gamma(B^{+}\rightarrow \bar{f}_{B^-})\right]\nonumber\\
&+2\left[\Gamma(B_c^{-}\rightarrow  B^- K_L^0)-\Gamma(B_c^{+}\rightarrow  B^+ K_L^0)\right]\cdot \Gamma(B^{-}\rightarrow f_{B^-})+\left[\left(\Gamma(B_c^{-}\rightarrow  B^- K_S^0)\right.\right.\nonumber\\
&\left.\left.-\Gamma(B_c^{+}\rightarrow  B^+ K_S^0)\right)-\left(\Gamma(B_c^{-}\rightarrow  B^- K_L^0)-\Gamma(B_c^{+}\rightarrow  B^+ K_L^0)\right)\right]\cdot  \Gamma(B^{-}\rightarrow f_{B^-})\nonumber\\
&=\left[\Gamma\left(B_c^{-}\rightarrow  B^- K_S^0\right)+\Gamma(B_c^{+}\rightarrow  B^+ K_S^0)+\Gamma(B_c^{-}\rightarrow  B^- K_L^0)+\Gamma(B_c^{+}\rightarrow  B^+ K_L^0)\right]\nonumber\\
&\cdot \Gamma(B^{-}\rightarrow f_{B^-})-\left[\Gamma(B_c^{+}\rightarrow  B^+ K_S^0)+\Gamma(B_c^{+}\rightarrow  B^+ K_L^0)\right]\cdot\left[\Gamma(B^{-}\rightarrow f_{B^-})-\Gamma(B^{+}\rightarrow \bar{f}_{B^-})\right].\label{Eq:cptasyobserre2}
\end{align}
By combining Eqs.(\ref{Eq:cptasyobserve1}), (\ref{Eq:cptasyobserve2}), (\ref{Eq:cptasyobserve3}), (\ref{Eq:cptasyobserre1}) and (\ref{Eq:cptasyobserre2}), we can obtain
\begin{align}
{\mathcal A}_{CPT}&\approx\frac{\Gamma(B_c^{+}\rightarrow  B^+ K_S^0)-\Gamma(B_c^{+}\rightarrow  B^+ K_L^0)}{\Gamma(B_c^{+}\rightarrow  B^+ K_S^0)+\Gamma(B_c^{+}\rightarrow  B^+ K_L^0)}\cdot\frac{\Gamma(B^{-}\rightarrow f_{B^-})-\Gamma(B^{+}\rightarrow \bar{f}_{B^-})}{\Gamma(B^{-}\rightarrow f_{B^-})+\Gamma(B^{+}\rightarrow \bar{f}_{B^-})}\nonumber\\
&-\frac{\Gamma(B_c^{+}\rightarrow  B^+ K_S^0)-\Gamma(B_c^{+}\rightarrow  B^+ K_L^0)}{\Gamma(B_c^{+}\rightarrow  B^+ K_S^0)+\Gamma(B_c^{+}\rightarrow  B^+ K_L^0)}\cdot\frac{\Gamma(B^{-}\rightarrow f_{B^-})-\Gamma(B^{+}\rightarrow \bar{f}_{B^-})}{\Gamma(B^{-}\rightarrow f_{B^-})+\Gamma(B^{+}\rightarrow \bar{f}_{B^-})}\nonumber\\
&\cdot\frac{\Gamma(B_c^{-}\rightarrow  B^-K_L^0)-\Gamma(B_c^{+}\rightarrow  B^+ K_L^0)}{\Gamma(B_c^{-}\rightarrow  B^-K_L^0)+\Gamma(B_c^{+}\rightarrow  B^+ K_L^0)}\nonumber\\
&-\frac{\Gamma(B_c^{+}\rightarrow  B^+ K_S^0)-\Gamma(B_c^{+}\rightarrow  B^+ K_L^0)}{\Gamma(B_c^{+}\rightarrow  B^+ K_S^0)+\Gamma(B_c^{+}\rightarrow  B^+ K_L^0)}\cdot\frac{\Gamma(B^{-}\rightarrow f_{B^-})-\Gamma(B^{+}\rightarrow \bar{f}_{B^-})}{\Gamma(B^{-}\rightarrow f_{B^-})+\Gamma(B^{+}\rightarrow \bar{f}_{B^-})}\nonumber\\
&\cdot\frac{\left[\Gamma(B_c^{-}\rightarrow  B^- K_S^0)-\Gamma(B_c^{+}\rightarrow  B^+ K_S^0)\right]-\left[\Gamma(B_c^{-}\rightarrow  B^- K_L^0)-\Gamma(B_c^{+}\rightarrow  B^+ K_L^0)\right]}{\Gamma(B_c^{-}\rightarrow  B^- K_S^0)+\Gamma(B_c^{+}\rightarrow  B^+ K_S^0)+\Gamma(B_c^{-}\rightarrow  B^- K_L^0)+\Gamma(B_c^{+}\rightarrow  B^+ K_L^0)}\nonumber\\
&+\frac{\left[\Gamma(B_c^{-}\rightarrow  B^- K_S^0)-\Gamma(B_c^{+}\rightarrow  B^+ K_S^0)\right]-\left[\Gamma(B_c^{-}\rightarrow  B^- K_L^0)-\Gamma(B_c^{+}\rightarrow  B^+ K_L^0)\right]}{\Gamma(B_c^{-}\rightarrow  B^- K_S^0)+\Gamma(B_c^{+}\rightarrow  B^+ K_S^0)+\Gamma(B_c^{-}\rightarrow  B^- K_L^0)+\Gamma(B_c^{+}\rightarrow  B^+ K_L^0)}\nonumber\\
&+\frac{\left[\Gamma(B_c^{-}\rightarrow  B^- K_S^0)-\Gamma(B_c^{+}\rightarrow  B^+ K_S^0)\right]-\left[\Gamma(B_c^{-}\rightarrow  B^- K_L^0)-\Gamma(B_c^{+}\rightarrow  B^+ K_L^0)\right]}{\Gamma(B_c^{-}\rightarrow  B^- K_S^0)+\Gamma(B_c^{+}\rightarrow  B^+ K_S^0)+\Gamma(B_c^{-}\rightarrow  B^- K_L^0)+\Gamma(B_c^{+}\rightarrow  B^+ K_L^0)}\nonumber\\
&\cdot\frac{\Gamma(B^{-}\rightarrow f_{B^-})-\Gamma(B^{+}\rightarrow \bar{f}_{B^-})}{\Gamma(B^{-}\rightarrow f_{B^-})+\Gamma(B^{+}\rightarrow \bar{f}_{B^-})},\label{Eq:cptasyobserrest}
\end{align}
then we obtain
\begin{align}
&{\mathcal A}_{CPT}=R\left(B_c^{+}\rightarrow  B^+ K_{S,L}^0 \right) \cdot{\mathcal A}_{CP}\left(B^-\rightarrow f_{B^-}\right)\nonumber\\
&-R\left(B_c^{+}\rightarrow  B^+ K_{S,L}^0 \right) \cdot{\mathcal A}_{CP}\left(B^-\rightarrow f_{B^-}\right)\cdot{\mathcal A}_{CP}\left(B_c^\pm\rightarrow B^{\pm}K_{L}^0\right)\nonumber\\
&-R\left(B_c^{+}\rightarrow  B^+ K_{S,L}^0 \right) \cdot{\mathcal A}_{CP}\left(B^-\rightarrow f_{B^-}\right)\cdot{\mathcal A}_{CPT}^{K_{S,L}^0}(B_c^{\pm}\rightarrow B^{\pm}K_{S,L}^0) \nonumber\\
&+{\mathcal A}_{CPT}^{K_{S,L}^0}(B_c^{\pm}\rightarrow B^{\pm}K_{S,L}^0)+{\mathcal A}_{CPT}^{K_{S,L}^0}(B_c^{\pm}\rightarrow B^{\pm}K_{S,L}^0) \cdot{\mathcal A}_{CP}\left(B^-\rightarrow f_{B^-}\right).\label{Eq:cptasyobserresh}
\end{align}
Neglecting the high order terms of $R\left(B_c^{+}\rightarrow  B^+ K_{S,L}^0 \right) \cdot{\mathcal A}_{CP}\left(B^-\rightarrow f_{B^-}\right)$ and ${\mathcal A}_{CPT}^{K_{S,L}^0}(B_c^{\pm}\rightarrow B^{\pm}K_{S,L}^0)$, we can obtain
\begin{align}
&{\mathcal A}_{CPT}=R\left(B_c^{+}\rightarrow  B^+ K_{S,L}^0 \right) \cdot{\mathcal A}_{CP}\left(B^-\rightarrow f_{B^-}\right)+{\mathcal A}_{CPT}^{K_{S,L}^0}(B_c^{\pm}\rightarrow B^{\pm}K_{S,L}^0)).\label{Eq:cptasyobsertot}
\end{align}
\end{appendix}


\begin{thebibliography}{100}

\bibitem{Sakharov:1967dj}
A.~D.~Sakharov,
Pisma Zh. Eksp. Teor. Fiz. \textbf{5}, 32-35 (1967).
doi:10.1070/PU1991v034n05ABEH002497.

\bibitem{Riotto:1998bt}
A.~Riotto,
[arXiv:hep-ph/9807454 [hep-ph]].

\bibitem{Buccella:1992sg}
F.~Buccella, M.~Lusignoli, G.~Mangano, G.~Miele, A.~Pugliese and P.~Santorelli,
Phys. Lett. B \textbf{302}, 319-325 (1993)
doi:10.1016/0370-2693(93)90402-4
[arXiv:hep-ph/9212253 [hep-ph]].

\bibitem{Buccella:1994nf}
F.~Buccella, M.~Lusignoli, G.~Miele, A.~Pugliese and P.~Santorelli,
Phys. Rev. D \textbf{51}, 3478-3486 (1995)
doi:10.1103/PhysRevD.51.3478
[arXiv:hep-ph/9411286 [hep-ph]].

\bibitem{Cheng:2012wr}
H.~Y.~Cheng and C.~W.~Chiang,
Phys. Rev. D \textbf{85}, 034036 (2012)
[erratum: Phys. Rev. D \textbf{85}, 079903 (2012)]
doi:10.1103/PhysRevD.85.034036
[arXiv:1201.0785 [hep-ph]].

\bibitem{Li:2021uhk}
Y.~B.~Li \textit{et al.} [Belle],
[arXiv:2103.06496 [hep-ex]].

\bibitem{Aaij:2020wil}
R.~Aaij \textit{et al.} [LHCb],
Eur. Phys. J. C \textbf{80}, no.10, 986 (2020)
doi:10.1140/epjc/s10052-020-8365-0
[arXiv:2006.03145 [hep-ex]].

\bibitem{Aaij:2019vnt}
R.~Aaij \textit{et al.} [LHCb],
Phys. Rev. Lett. \textbf{122}, no.19, 191803 (2019)
doi:10.1103/PhysRevLett.122.191803
[arXiv:1903.01150 [hep-ex]].

\bibitem{Li:2021iwf}
H.~B.~Li and X.~R.~Lyu,
[arXiv:2103.00908 [hep-ex]].

\bibitem{Zhang:2021zhr}
Z.~H.~Zhang,
[arXiv:2102.12263 [hep-ph]].

\bibitem{Lenz:2020awd}
A.~Lenz and G.~Wilkinson,
doi:10.1146/annurev-nucl-102419-124613
[arXiv:2011.04443 [hep-ph]].

\bibitem{Unal:2020ezc}
Y.~\"Unal and U.~G.~Mei\ss{}ner,
JHEP \textbf{01}, 115 (2021)
doi:10.1007/JHEP01(2021)115
[arXiv:2008.01371 [hep-ph]].

\bibitem{Saur:2020rgd}
M.~Saur and F.~S.~Yu,
Sci. Bull. \textbf{65}, 1428-1431 (2020)
doi:10.1016/j.scib.2020.04.020
[arXiv:2002.12088 [hep-ex]].

\bibitem{Wang:2017gxe}
D.~Wang, P.~F.~Guo, W.~H.~Long and F.~S.~Yu,
JHEP \textbf{03}, 066 (2018)
doi:10.1007/JHEP03(2018)066
[arXiv:1709.09873 [hep-ph]].

\bibitem{Yu:2017oky}
D.~Wang, F.~S.~Yu and H.~n.~Li,
Phys. Rev. Lett. \textbf{119}, no.18, 181802 (2017)
doi:10.1103/PhysRevLett.119.181802
[arXiv:1707.09297 [hep-ph]].

\bibitem{Cheng:2012xb}
H.~Y.~Cheng and C.~W.~Chiang,
Phys. Rev. D \textbf{86}, 014014 (2012)
doi:10.1103/PhysRevD.86.014014
[arXiv:1205.0580 [hep-ph]].

\bibitem{Azimov:1999gw}
Y.~I.~Azimov,
[arXiv:hep-ph/9910368 [hep-ph]].

\bibitem{Azimov:1999dj}
Y.~I.~Azimov,
[arXiv:hep-ph/9907260 [hep-ph]].

\bibitem{Lipkin:1999qz}
H.~J.~Lipkin and Z.~z.~Xing,
Phys. Lett. B \textbf{450}, 405-411 (1999)
doi:10.1016/S0370-2693(99)00170-7
[arXiv:hep-ph/9901329 [hep-ph]].

\bibitem{Xing:1995jg}
Z.~Z.~Xing,
Phys. Lett. B \textbf{353}, 313-318 (1995)
[erratum: Phys. Lett. B \textbf{363}, 266 (1995)]
doi:10.1016/0370-2693(95)92845-D
[arXiv:hep-ph/9505272 [hep-ph]].

\bibitem{Aaij:2019kcg}
R.~Aaij \textit{et al.} [LHCb],
Phys. Rev. Lett. \textbf{122}, no.21, 211803 (2019)
doi:10.1103/PhysRevLett.122.211803
[arXiv:1903.08726 [hep-ex]].

\bibitem{Lueders:1992dq}
G.~Luders,
Annals Phys. \textbf{2}, 1-15 (1957)
doi:10.1016/0003-4916(57)90032-5.

\bibitem{Karan:2017coa}
A.~Karan, A.~K.~Nayak, R.~Sinha and D.~London,
Phys. Lett. B \textbf{781}, 459-463 (2018)
doi:10.1016/j.physletb.2018.04.029
[arXiv:1712.01298 [hep-ph]].

\bibitem{Anastasi:2018qqf}
A.~Anastasi \textit{et al.} [KLOE-2],
JHEP \textbf{09}, 021 (2018)
doi:10.1007/JHEP09(2018)021
[arXiv:1806.08654 [hep-ex]].

\bibitem{Karan:2020ada}
A.~Karan and A.~K.~Nayak,
Phys. Rev. D \textbf{101}, no.1, 015027 (2020)
doi:10.1103/PhysRevD.101.015027
[arXiv:2001.05282 [hep-ph]].

\bibitem{Karan:2020yhk}
A.~Karan,
Eur. Phys. J. C \textbf{80}, no.8, 782 (2020)
doi:10.1140/epjc/s10052-020-8297-8
[arXiv:2007.06725 [hep-ph]].

\bibitem{Domenico:2020bbk}
A.~Di Domenico,
Symmetry \textbf{12}, no.12, 2063 (2020)
doi:10.3390/sym12122063.

\bibitem{Amorim:1998pi}
A.~Amorim, M.~G.~Santos and J.~P.~Silva,
Phys. Rev. D \textbf{59}, 056001 (1999)
doi:10.1103/PhysRevD.59.056001
[arXiv:hep-ph/9807364 [hep-ph]].

\bibitem{Ko:2012pe}
  B.~R.~Ko {\it et al.} [Belle Collaboration],
  Phys.\ Rev.\ Lett.\  {\bf 109}, 021601 (2012)
  Erratum: [Phys.\ Rev.\ Lett.\  {\bf 109}, 119903 (2012)]
  doi:10.1103/PhysRevLett.109.021601, 10.1103/PhysRevLett.109.119903
  [arXiv:1203.6409 [hep-ex]].

\bibitem{Ko:2010ng}
B.~R.~Ko \textit{et al.} [Belle],
Phys. Rev. Lett. \textbf{104}, 181602 (2010)
doi:10.1103/PhysRevLett.104.181602
[arXiv:1001.3202 [hep-ex]].

\bibitem{delAmoSanchez:2011zza}
P.~del Amo Sanchez \textit{et al.} [BaBar],
Phys. Rev. D \textbf{83}, 071103 (2011)
doi:10.1103/PhysRevD.83.071103
[arXiv:1011.5477 [hep-ex]].

\bibitem{BABAR:2011aa}
J.~P.~Lees \textit{et al.} [BaBar],
Phys. Rev. D \textbf{85}, 031102 (2012)
[erratum: Phys. Rev. D \textbf{85}, 099904 (2012)]
doi:10.1103/PhysRevD.85.031102
[arXiv:1109.1527 [hep-ex]].

\bibitem{Mendez:2009aa}
H.~Mendez \textit{et al.} [CLEO],
Phys. Rev. D \textbf{81}, 052013 (2010)
doi:10.1103/PhysRevD.81.052013
[arXiv:0906.3198 [hep-ex]].

\bibitem{Dobbs:2007ab}
S.~Dobbs \textit{et al.} [CLEO],
Phys. Rev. D \textbf{76}, 112001 (2007)
doi:10.1103/PhysRevD.76.112001
[arXiv:0709.3783 [hep-ex]].

\bibitem{Link:2001zj}
J.~M.~Link \textit{et al.} [FOCUS],
Phys. Rev. Lett. \textbf{88}, 041602 (2002)
[erratum: Phys. Rev. Lett. \textbf{88}, 159903 (2002)]
doi:10.1103/PhysRevLett.88.041602
[arXiv:hep-ex/0109022 [hep-ex]].

\bibitem{Grossman:2011zk}
Y.~Grossman and Y.~Nir,
JHEP \textbf{04}, 002 (2012)
doi:10.1007/JHEP04(2012)002
[arXiv:1110.3790 [hep-ph]].

\bibitem{Bigi:2012km}
I.~I.~Bigi,
[arXiv:1204.5817 [hep-ph]].

\bibitem{Poireau:2012by}
V.~Poireau [BaBar],
[arXiv:1205.2201 [hep-ex]].

\bibitem{Chen:2020uxi}
F.~Z.~Chen, X.~Q.~Li and Y.~D.~Yang,
JHEP \textbf{05}, 151 (2020)
doi:10.1007/JHEP05(2020)151
[arXiv:2003.05735 [hep-ph]];
F.~Z.~Chen, X.~Q.~Li, Y.~D.~Yang and X.~Zhang,
Phys. Rev. D \textbf{100}, no.11, 113006 (2019)
doi:10.1103/PhysRevD.100.113006
[arXiv:1909.05543 [hep-ph]];
A.~Dighe, S.~Ghosh, G.~Kumar and T.~S.~Roy,
[arXiv:1902.09561 [hep-ph]];
J.~Rend\'on, P.~Roig and G.~Toledo,
Phys. Rev. D \textbf{99}, no.9, 093005 (2019)
doi:10.1103/PhysRevD.99.093005
[arXiv:1902.08143 [hep-ph]];
V.~Cirigliano, A.~Crivellin and M.~Hoferichter,
SciPost Phys. Proc. \textbf{1}, 007 (2019)
doi:10.21468/SciPostPhysProc.1.007;%
G.~L\'opez Castro,
SciPost Phys. Proc. \textbf{1}, 008 (2019)
doi:10.21468/SciPostPhysProc.1.008
[arXiv:1812.05892 [hep-ph]];
%
D.~Delepine, G.~Faisel and C.~A.~Ramirez,
[arXiv:1806.05090 [hep-ph]];
V.~Cirigliano, A.~Crivellin and M.~Hoferichter,
Phys. Rev. Lett. \textbf{120}, no.14, 141803 (2018)
doi:10.1103/PhysRevLett.120.141803
[arXiv:1712.06595 [hep-ph]];
L.~Dhargyal,
LHEP \textbf{1}, no.3, 9-14 (2018)
doi:10.31526/LHEP.3.2018.03
[arXiv:1605.00629 [hep-ph]];
H.~Z.~Devi, L.~Dhargyal and N.~Sinha,
Phys. Rev. D \textbf{90}, no.1, 013016 (2014)
doi:10.1103/PhysRevD.90.013016
[arXiv:1308.4383 [hep-ph]];
D.~Kimura, K.~Y.~Lee and T.~Morozumi,
PTEP \textbf{2013}, 053B03 (2013)
[erratum: PTEP \textbf{2013}, no.9, 099201 (2013); erratum: PTEP \textbf{2014}, no.8, 089202 (2014)]
doi:10.1093/ptep/ptt013
[arXiv:1201.1794 [hep-ph]].

\bibitem{Brambilla:2010cs}
N.~Brambilla, S.~Eidelman, B.~K.~Heltsley, R.~Vogt, G.~T.~Bodwin, E.~Eichten, A.~D.~Frawley, A.~B.~Meyer, R.~E.~Mitchell and V.~Papadimitriou, \textit{et al.}
Eur. Phys. J. C \textbf{71}, 1534 (2011)
doi:10.1140/epjc/s10052-010-1534-9
[arXiv:1010.5827 [hep-ph]].

\bibitem{QuarkoniumWorkingGroup:2004kpm}
N.~Brambilla \textit{et al.} [Quarkonium Working Group],
doi:10.5170/CERN-2005-005
[arXiv:hep-ph/0412158 [hep-ph]].

\bibitem{Zheng:2020ult}
T.~Zheng, J.~Xu, L.~Cao, D.~Yu, W.~Wang, S.~Prell, Y.~K.~E.~Cheung and M.~Ruan,
Chin. Phys. C \textbf{45}, no.2, 023001 (2021)
doi:10.1088/1674-1137/abcf1f
[arXiv:2007.08234 [hep-ex]].

\bibitem{Cheng:2021svx}
W.~Cheng, Y.~Zhang, L.~Zeng, H.~B.~Fu and X.~G.~Wu,
[arXiv:2107.08405 [hep-ph]].

\bibitem{Zhou:2020bnm}
T.~Zhou, T.~Wang, H.~F.~Fu, Z.~H.~Wang, L.~Huo and G.~L.~Wang,
Eur. Phys. J. C \textbf{81}, no.4, 339 (2021)
doi:10.1140/epjc/s10052-021-09128-2
[arXiv:2012.06135 [hep-ph]].

\bibitem{Xiao:2011zz}
Z.~J.~Xiao and X.~Liu,
Phys. Rev. D \textbf{84}, 074033 (2011)
doi:10.1103/PhysRevD.84.074033
[arXiv:1111.6679 [hep-ph]].

\bibitem{Fakirov:1977ta}
  D.~Fakirov and B.~Stech,
  Nucl.\ Phys.\ B {\bf 133} (1978) 315.
  doi:10.1016/0550-3213(78)90306-1.

\bibitem{Wang:2008tf}
  Y.~M.~Wang, H.~Zou, Z.~T.~Wei, X.~Q.~Li and C.~D.~Lu,
  Eur.\ Phys.\ J.\ C {\bf 55} (2008) 607
  doi:10.1140/epjc/s10052-008-0619-1
  [arXiv:0802.2762 [hep-ph]].

\bibitem{Giri:1998qf}
  A.~K.~Giri and R.~Mohanta,
  Pramana {\bf 50} (1998) 155.
  doi:10.1007/BF02847526.

\bibitem{Branco:1999fs}
  G.~C.~Branco, L.~Lavoura and J.~P.~Silva,
  Int.\ Ser.\ Monogr.\ Phys.\  {\bf 103}, 1 (1999).

\bibitem{Cooper:2020wnj}
  L.~J.~Cooper, C.~T.~H.~Davies, J.~Harrison, J.~Komijani, M.~Wingate [HPQCD Collaboration],
  Phys.\ Rev.\ D {\bf 102}, no. 1, 014513 (2020)
  doi:10.1103/PhysRevD.102.014513
  [arXiv:2003.00914 [hep-lat]].

\bibitem{Zyla:2020zbs}
P.A.~Zyla \textit{et al.} [Particle Data Group],
PTEP \textbf{2020}, no.8, 083C01 (2020)
doi:10.1093/ptep/ptaa104.

\bibitem{Cerri:2018ypt}
A.~Cerri, V.~V.~Gligorov, S.~Malvezzi, J.~Martin Camalich, J.~Zupan, S.~Akar, J.~Alimena, B.~C.~Allanach, W.~Altmannshofer and L.~Anderlini, \textit{et al.}
CERN Yellow Rep. Monogr. \textbf{7}, 867-1158 (2019)
doi:10.23731/CYRM-2019-007.867
[arXiv:1812.07638 [hep-ph]].

\bibitem{D0:2016xvr}
V.~M.~Abazov \textit{et al.} [D0],
Phys. Rev. D \textbf{95}, no.3, 031101 (2017)
doi:10.1103/PhysRevD.95.031101
[arXiv:1608.00863 [hep-ex]].

\bibitem{Bigi:1994aw}
I.~I.~Y.~Bigi and H.~Yamamoto,
Phys. Lett. B \textbf{349}, 363-366 (1995)
doi:10.1016/0370-2693(95)00285-S
[arXiv:hep-ph/9502238 [hep-ph]].

\bibitem{Wang:2017ksn}
D.~Wang, F.~S.~Yu, P.~F.~Guo and H.~Y.~Jiang,
Phys. Rev. D \textbf{95}, no.7, 073007 (2017)
doi:10.1103/PhysRevD.95.073007
[arXiv:1701.07173 [hep-ph]].

\bibitem{CLEO:2007rhw}
Q.~He \textit{et al.} [CLEO],
Phys. Rev. Lett. \textbf{100}, 091801 (2008)
doi:10.1103/PhysRevLett.100.091801
[arXiv:0711.1463 [hep-ex]].

\bibitem{Bailey:2018feb}
J.~A.~Bailey, S.~Lee, W.~Lee, J.~Leem and S.~Park,
Phys. Rev. D \textbf{98}, no.9, 094505 (2018)
doi:10.1103/PhysRevD.98.094505
[arXiv:1808.09657 [hep-lat]].

\bibitem{ExtendedTwistedMass:2020tvp}
G.~Bergner \textit{et al.} [Extended Twisted Mass],
PoS \textbf{LATTICE2019}, 181 (2020)
doi:10.22323/1.363.0181
[arXiv:2001.09116 [hep-lat]].

\bibitem{Buras:1998raa}
A.~J.~Buras,
[arXiv:hep-ph/9806471 [hep-ph]].

\bibitem{Ref:utfit}
  [UTfit Collaboration],
  online update at:\\ http://utfit.org/UTfit/ResultsSummer2018SM.

\bibitem{Bianco:2003vb}
S.~Bianco, F.~L.~Fabbri, D.~Benson and I.~Bigi,
Riv. Nuovo Cim. \textbf{26}, no.7-8, 1-200 (2003)
doi:10.1393/ncr/i2003-10003-1
[arXiv:hep-ex/0309021 [hep-ex]].

\bibitem{Petrov:2011un}
A.~A.~Petrov,
[arXiv:1101.3822 [hep-ph]].

\bibitem{LHCb:2013zpr}
R.~Aaij \textit{et al.} [LHCb],
Phys. Rev. Lett. \textbf{111}, no.25, 251801 (2013)
doi:10.1103/PhysRevLett.111.251801
[arXiv:1309.6534 [hep-ex]].

\bibitem{LHCb:2016qsa}
R.~Aaij \textit{et al.} [LHCb],
Phys. Rev. D \textbf{95}, no.5, 052004 (2017)
[erratum: Phys. Rev. D \textbf{96}, no.9, 099907 (2017)]
doi:10.1103/PhysRevD.95.052004
[arXiv:1611.06143 [hep-ex]].


\bibitem{LHCb:2018roe}
R.~Aaij \textit{et al.} [LHCb],
[arXiv:1808.08865 [hep-ex]].

\bibitem{LHCb:2012myk}
R.~Aaij \textit{et al.} [LHCb],
Eur. Phys. J. C \textbf{73}, no.4, 2373 (2013)
doi:10.1140/epjc/s10052-013-2373-2
[arXiv:1208.3355 [hep-ex]].

\bibitem{Dai:1998hb}
Y.~S.~Dai and D.~S.~Du,
Eur. Phys. J. C \textbf{9}, 557-564 (1999)
doi:10.1007/s100529900073
[arXiv:hep-ph/9809386 [hep-ph]].

\bibitem{Fu:2011tn}
H.~F.~Fu, Y.~Jiang, C.~S.~Kim and G.~L.~Wang,
JHEP \textbf{06}, 015 (2011)
doi:10.1007/JHEP06(2011)015
[arXiv:1102.5399 [hep-ph]].

\end{thebibliography}
\end{document}